\documentclass[%
 reprint,
 amsmath,amssymb,
 aps,
prb,
floatfix,
]{revtex4-2}

\usepackage{placeins}

\usepackage{xcolor}
\usepackage[normalem]{ulem}
\usepackage{comment}
\usepackage{graphicx}
\usepackage{dcolumn}
\usepackage{bm}
\usepackage{multirow}
\usepackage{hyperref}
\hypersetup{colorlinks=true,breaklinks,linkcolor=blue,urlcolor=blue,citecolor=blue}

\newcommand{\TUVienna}{\affiliation{$^a$Institute of Solid State Physics, TU Wien, 1040 Vienna, Austria}}
\newcommand{\LMUMunich}{\affiliation{$^b$Arnold Sommerfeld Center for Theoretical Physics, Center for NanoScience, and Munich Center for Quantum Science and Technology, Ludwig-Maximilians-Universit\"at M\"unchen, 80333 Munich, Germany}}

\begin{document}

\preprint{}

\title{Highly nonperturbative nature of the Mott metal-insulator transition:\\Two-particle vertex divergences in the coexistence region}

\author{M. Pelz$^{a,b}$}
\author{S. Adler$^{a}$}
\author{M. Reitner$^{a}$}
\author{A. Toschi$^{a}$}

\TUVienna
\LMUMunich

\date{\today}

\begin{abstract}
We thoroughly analyze the divergences of the irreducible vertex functions occurring in the charge channel of the half-filled Hubbard model in close proximity to the Mott metal-insulator transition (MIT). In particular, by systematically performing  dynamical mean-field theory (DMFT) calculations on the two-particle level, we determine the location and the number of the vertex divergences across the whole coexistence region adjacent to the first-order metal-to-insulator transition. We find that the lines in the parameter space, along which the vertex divergences occur, display a qualitatively different shape in the coexisting metallic and insulating phase, which is also associated to an abrupt jump of the number of divergences across the MIT. Physically, the systematically larger number of divergences on the insulating side of the transition reflects the sudden suppression of local charge fluctuation at the MIT. Further, a systematic analysis of the results demonstrates that the number of divergence lines increases as a function of the inverse temperature ${\beta\!=\!(k_\mathrm{B} T)^{-1}}$ by approaching the Mott transition in the zero temperature limit. This makes it possible to identify the zero-temperature MIT as an \textit{accumulation} point of an \textit{infinite} number of vertex divergence lines, unveiling the highly nonperturbative nature of the underlying transition.
\end{abstract}

\maketitle

\section{\label{Introduction}Introduction}
The multifaceted manifestations \cite{Schaefer2013,Janis2014,Kozik2015,Stan2015,Rossi2015,Ribic2016,Rossi2016,Gunnarsson2016,Schaefer2016,Gunnarsson2017} of the breakdown of the self-consistent perturbation theory for the many-electron problem have been recently in focus of several studies \cite{Tarantino2018,Vucicevic2018,Chalupa2018,Thunstroem2018,Nourafkan2019,Melnick2020,Springer2020,Kim2020,vanLoon2020,Reitner2020,Chalupa2021,Kozik2021,Mazitov2022,Stepanov2022,Mazitov2022b,vanLoon2022,Adler2022,Kim2022Kozik}. 
In particular, it has been demonstrated \cite{Gunnarsson2017} how the breakdown of the perturbative expansion corresponds to the crossings of different solutions \cite{Kozik2015,Stan2015} of the Luttinger-Ward functional or, equivalently, to multiple divergences \cite{Schaefer2013,Schaefer2016,Chalupa2018} of the two-particle vertex functions irreducible in the charge channel, i.e.~the kernel of the Bethe-Salpeter equation describing the charge response of the many-electron system under consideration.

On a less formal perspective, the crucial role \cite{Chalupa2021,Mazitov2022,Mazitov2022b,Adler2022} of the local moment formation in triggering the perturbation-theory breakdown as well as the contra-intuitive physical consequences associated to nonperturbative scattering processes \cite{Reitner2020} have been extensively investigated in the most recent literature, e.g.~for the Anderson impurity and the Hubbard model.

Hitherto, however, the analysis of an equally important aspect of this problem, namely the precise relation linking the above mentioned manifestations of the perturbative breakdown to the occurrence of Mott-Hubbard metal-to-insulator transitions (MITs) \cite{Imada1998}, has been put, to some extent, aside.
In fact, in some of the earliest works \cite{Schaefer2013,Schaefer2016} on this subject, it was suggested that the divergences of the irreducible vertex could be viewed as ``precursors" of the Mott-Hubbard MIT. In later studies, however, it was shown \cite{Chalupa2018} that multiple irreducible vertex divergences were also occurring in cases, such as the Anderson Impurity model (AIM), where {\sl no} Mott-Hubbard transition takes place. The  contradiction of this observation with the previously proposed interpretation has left the full understanding of this aspect of the perturbative breakdown unsolved.

In this work, we will hence address the still outstanding question of how the Mott MIT, which represents an intrinsically nonperturbative phenomenon, is actually related to the breakdown of the perturbation expansion by analyzing one of its characterizing manifestations, the divergences of the irreducible vertex functions.

To this aim, we will perform systematic dynamical mean-field theory (DMFT) \cite{Georges1996} calculations of the two-particle Green's/vertex functions \cite{Rohringer2012} of the Hubbard model in one of its most delicate parameter regimes, the coexistence region of the Mott MIT, which was not considered in preceding studies on this specific topic.

In particular, we will determine the location, the number, and the properties of the vertex divergences occurring in both the (coexisting) metallic and insulating DMFT solution of the half-filled Hubbard model in the proximity of the Mott transition. Hereby, we focus on the changes taking place across the finite-$T$ first-order transition and on the extrapolation of the corresponding results towards the ${T\!=\!0}$ second-order (quantum) critical endpoint of the MIT. This procedure will eventually allow to draw rigorous conclusions about the link between vertex divergences and the Mott-Hubbard MIT.

The plan of the paper is the following: In Sec.~II, we introduce the model and the formalism necessary for our analysis and briefly recapitulate  results of previous studies relevant for our scopes; in Sec.~III we illustrate our DMFT results for the divergences of the irreducible vertex functions systematically obtained in the coexistence region of the Mott MIT and the corresponding extrapolation performed down to zero-temperature; then in Sec.~IV we will discuss the overall scenario emerging from our study, and, eventually, in Sec.~V we will present the conclusions and the outlook of our work.

\section{\label{Model_Fromalism}Model,  Formalism and Methods}

\subsection{\label{Model}Model}
In this study we consider a single-band Hubbard model (HM) \cite{Hubbard1963} on the Bethe lattice with infinite connectivity, whose density of state is semicircular with half-bandwidth ${D\!=\!1}$, which serves as unit of energy throughout the work.

The Hamiltonian reads
\begin{equation}
    \label{eq:HM}
    \mathcal{H} = -t\sum_{\left<i\,j\right>\,\sigma}c^{\dagger}_{i\sigma}c_{j\sigma} +  U\sum_{i}n_{i\uparrow}n_{i\downarrow}\,,
\end{equation}
where ${t\!=\!\frac{1}{2D}}$ is the (spin-independent) nearest neighbor hopping between neighboring lattice sites $i$ and $j$, and ${c^{\dagger}_{i\sigma}}$, (${c_{i\sigma}}$) the fermionic creation (annihilation) operator with spin ${\sigma\!=\,\uparrow,\downarrow}$ at site $i$, and $U$ is the local (Hubbard) repulsive interaction between two electrons on the same lattice site (${n_{i\sigma}\!=\!c^{\dagger}_{i\sigma}c_{i\sigma}}$ denoting the particle-number operator at site $i$ for spin $\sigma$).

\begin{figure*}[t!]
    \centering
    \includegraphics[width=\linewidth]{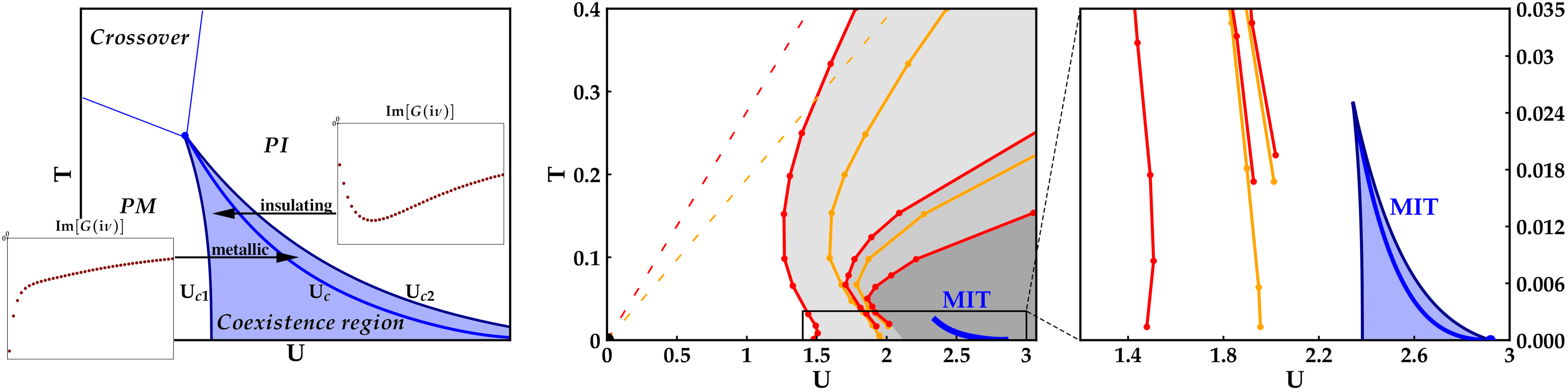}
    \caption{Location of the irreducible vertex divergence lines in the phase-diagram of Hubbard model on a square lattice solved by dynamical mean field theory (DMFT). Left panel: Schematic illustration of the MIT and its coexistence region in DMFT. The insets show examples of the imaginary part of the one-particle Green's function on the Matsubara frequency axis for the PM (left inset) and the PI (right inset) DMFT solution, featuring completely different low-energy behaviors. The arrows represent the scanning direction for the respective convergent solution of the two phases in the coexistence region. Middle panel: Reproduced from Ref.~\cite{Schaefer2016}, solid red and orange lines \cite{color_coding} mark the first five $\boldsymbol{\Gamma}_c^\infty$-lines of the HM bending around the metal-insulator transition (blue solid line). Red and orange dashed lines show the first two $\boldsymbol{\Gamma}_c^\infty$-lines of the Hubbard atom (HA) for comparison. Right panel: Close-up of the region near the MIT marked in the middle panel. The blue shaded area indicates the coexistence region of the MIT. The thermodynamic phase transition ($U_c$) is marked in blue and the borders of the coexistence region ($U_{c1}$ and $U_{c2}$) are displayed as dark blue lines. To the left in red and orange the closest vertex divergence lines from \cite{Schaefer2016} are visible.}
    \label{fig:intro}
\end{figure*}

We set the density to half-filling ${\big(\big<n_\uparrow\big>\!=\!\big<n_\downarrow\big>\!=\!\frac{1}{2}\big)}$, where the HM we consider, which can be exactly solved by means of dynamical mean-field theory (DMFT), 
is known to feature a paradigmatic realization of the Mott-Hubbard MIT transition.
Specifically, we briefly recall here \cite{Georges1996} that, by neglecting possible SU(2) symmetry-broken phases, the DMFT solution of Eq.~(\ref{eq:HM}) yields a first-order transition between a paramagnetic metallic (PM) and a paramagnetic insulating phase (PI) along a $U_c(T)$ transition line, ending with second-order critical points at finite ${T\!=\!T_c\!\approx\!\frac{1}{39}}$ and at ${T\!=\!0}$, as schematically shown in the leftmost panel of Fig.~\ref{fig:intro}. The first-order nature of the transition is witnessed by the presence of a broad hysteresis region, delimited by the lines $U_{c_1}(T)$ (${<\!U_c(T)}$ on the left side) and $U_{c_2}(T)$ (${>\!U_c(T)}$ on the right side) where a PM and a PI numerical DMFT solution coexists \cite{DMFT_boarder}, with distinct physical properties (as exemplified by the qualitatively different shape of the corresponding one particle Green's functions, shown in the inset of the leftmost panel in Fig.~\ref{fig:intro}).

\subsection{\label{Formalism}Formalism}
As mentioned in the Introduction, an evident manifestation of the breakdown of the self-consistent perturbation expansion in many-electron problems is the divergence of the kernel of the Bethe-Salpeter equation (BSE) for the system response in the charge sector. Hence, the central quantity for our DMFT investigation will be the on-site generalized susceptibility, which, in the imaginary time-domain, is defined as:
\begin{align}
	&\boldsymbol{\chi}_{\sigma_1,\sigma_2,\sigma_3,\sigma_4}(\tau_1,\tau_2,\tau_3,\tau_4):=&\nonumber\\ &\hspace{1cm}\bigg[\big<\mathcal{T}_\tau c^\dagger_{\sigma_1}(\tau_1) c_{\sigma_2}(\tau_2) c^\dagger_{\sigma_3}(\tau_3) c_{\sigma_4}(\tau_4)\big>\,-&\nonumber\\&\hspace{2cm}\big<\mathcal{T}_\tau c^\dagger_{\sigma_1}(\tau_1) c_{\sigma_2}(\tau_2)\big>\big<\mathcal{T}_\tau c^\dagger_{\sigma_3}(\tau_3) c_{\sigma_4}(\tau_4)\big>\bigg]
	\\& \hspace{1cm} := G^{(2)}_{\sigma_1,\sigma_2,\sigma_3,\sigma_4}(\tau_1,\tau_2,\tau_3,\tau_4)\,-&\nonumber\\&\hspace{2cm}G_{\sigma_1,\sigma_2}(\tau_1,\tau_2)\,G_{\sigma_3,\sigma_4}(\tau_3,\tau_4),&
	\label{eq:chitau}
\end{align}
in terms of the one ${G_{\sigma_1,\sigma_2}(\tau_1,\tau_2)}$ and two-particle ${G^{(2)}_{\sigma_1,\sigma_2,\sigma_3,\sigma_4}(\tau_1,\tau_2,\tau_3,\tau_4)}$ local Green's functions of the DMFT solution, where $\sigma_i$ denotes the spins of the incoming/outgoing particles and $\tau_i$ their imaginary time arguments \cite{Rohringer2012}. 
By taking the Fourier transform into Matsubara frequencies, and exploiting the time-translational invariance as well as the SU(2) symmetry of the problem, one then obtains \cite{Rohringer2012}
\begin{align}
	&\boldsymbol{\chi}_{ph,\sigma\sigma^\prime}^{\nu\nu^\prime\Omega}=\int_{0}^{\beta}d\tau_1d\tau_2d\tau_3\,e^{-i\nu\tau_1}e^{i(\nu+\Omega)\tau_2}e^{-i(\nu^\prime+\Omega)\tau_3}\,\cdot &\nonumber\\&\hspace{15mm} \boldsymbol{\chi}_{\sigma,\sigma,\sigma',\sigma'}(\tau_1,\tau_2,\tau_3,0),
	 \label{eq:chiph}
  \end{align}
where, in the particle-hole convention we adopted, $\nu$, $\nu'$ ($\Omega$) denote fermionic (bosonic) Matsubara frequencies, respectively.
We recall that the expression in Eq.~(\ref{eq:chiph}) can be directly connected to the physical charge response $\chi_c(\Omega)$ of the system by performing a summation over both fermionic Matsubara frequencies $\nu$, $\nu'$ and spin indices:
\begin{equation}
 \chi_c(\Omega) = \frac{1}{\beta^2} \, \sum_{\nu, \nu'}   \left(\boldsymbol{\chi}_{ph,\uparrow\uparrow}^{\nu\nu'\Omega}+\boldsymbol{\chi}_{ph,\uparrow\downarrow}^{\nu\nu'\Omega} \right).
 \label{eq:chiphys}
\end{equation}

\begin{figure}[h!]
\includegraphics[width=\linewidth]{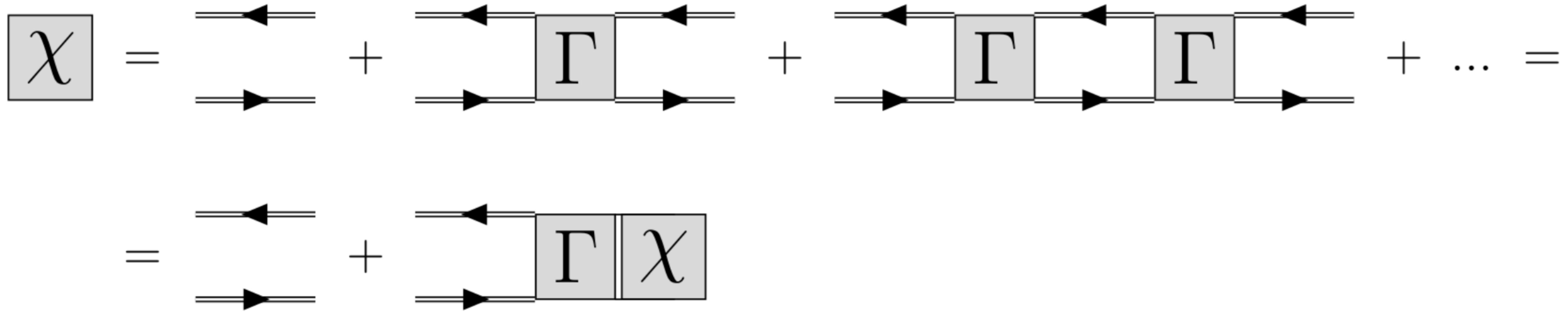}
\caption{\label{eq:BSE} Diagrammatic representation of the Bethe-Salpeter equation (BSE) ${\boldsymbol{\chi}\!=\!\boldsymbol{\chi}_0\!-\!\boldsymbol{\chi}_0\boldsymbol{\Gamma}\boldsymbol{\chi}}$, where $\boldsymbol{\chi}$ is the generalized susceptibility, $\boldsymbol{\chi}_0$ the bubble term ${\bigl(\propto\!-\!\beta G(\nu)\,G(\nu\!+\!\Omega)\,\delta_{\nu\nu^\prime}\bigr)}$, and $\boldsymbol{\Gamma}$ contains all irreducible vertex diagrams \cite{Schaefer2016,Nakanishi1969}.}
\end{figure}

Consistent with the SU(2) symmetry of the problem, the BSE of the generalized response in the charge sector ${\boldsymbol{\chi}_c^{\nu\nu'\Omega}\!=\!\boldsymbol{\chi}_{ph,\uparrow\uparrow}^{\nu\nu'\Omega}\!+\!\boldsymbol{\chi}_{ph,\uparrow\downarrow}^{\nu\nu'\Omega}}$ can be recasted, for each value of the bosonic frequency $\Omega$, in a closed matrix equation for the fermionic Matsubara frequency space, as diagrammatically shown in Fig.~\ref{eq:BSE}.

The equation describes the infinitely repeated insertions of (two-particle) irreducible vertex corrections (${\boldsymbol{\Gamma}_c}$) to the independent propagation of a particle and a hole (i.e., the so-called bubble term: ${\boldsymbol{\chi}_{c,0}^{\nu\nu',\Omega}\!=\!-\beta\,G(\nu)\,G(\nu\!+\!\Omega)\,\delta_{\nu\nu'}}$, where $G(\nu)$ denotes the one-particle local DMFT Green's function on the Matsubara axis).  
Hence, the irreducible vertex function, which represents the kernel of the BSE, is defined (and can be explicitly computed) as \cite{Schaefer2016}
\begin{equation}
    \label{eq:inverseBSE}
    \boldsymbol{\Gamma}_c^{\nu\nu'\Omega}=\beta^2 \bigg(\big[\boldsymbol{\chi}_{c}^\Omega\big]_{\nu\nu'}^{-1}-\big[\boldsymbol{\chi}_{c,0}^\Omega\big]_{\nu\nu'}^{-1}\bigg).
\end{equation}

In this work, we consider the case of zero bosonic frequency, which is linked, according to Eq.~(\ref{eq:chiphys}), to the isothermal (or static) charge response \cite{Wilcox1968,Watzenboeck2022} of the system, as the corresponding vertex divergences are the first to occur by increasing values of $U$ and are those directly related \cite{Gunnarsson2017} to the crossing of solutions of the Luttinger-Ward functional. Hence, for the sake of readability, we will denote ${\boldsymbol{\Gamma}_c^{\nu\nu' \Omega=0}\equiv\boldsymbol{\Gamma}_c^{\nu \nu'}}$ and ${\boldsymbol{\chi}_c^{\nu\nu' \Omega=0}\equiv\boldsymbol{\chi}_c^{\nu \nu'}}$.

By a quick inspection of Eq.~(\ref{eq:inverseBSE}), one understands that, at any finite temperature, no divergence of $\boldsymbol{\Gamma}_c$ can be caused by the inversion of the (frequency diagonal) bubble term, as ${G(\nu)\!\neq\!0}$ for all $\nu$. 
As a consequence, all divergences of $\boldsymbol{\Gamma}_c$ must originate from the non-invertibility of the generalized susceptibility matrix or, more formally, from the vanishing of one of its eigenvalues $\lambda_i$ \cite{Schaefer2013,Schaefer2016}.
Hence, all points $(T,U)$, where a specific eigenvalue $\lambda_i$ of $\boldsymbol{\chi}_c$ crosses zero, define a line in the phase-diagram of the half-filled HM, where the corresponding irreducible charge vertex diverges ($\boldsymbol{\Gamma}_c^\infty$-line). It is important to stress, here, that since for small $U$, ${\boldsymbol{\chi}_c\!\approx\!\boldsymbol{\chi}_{c,0}}$ , and the latter is a positive definite (diagonal) matrix, the number of {\sl negative} eigenvalues ($N_{\lambda<0}$) of the generalized charge susceptibility matrix at a specific parameter set corresponds to the number of crossed $\boldsymbol{\Gamma}_c^\infty$-lines coming from $U\!=\!0$ and can be also used to approximate the shape of the $\boldsymbol{\Gamma}_c^\infty$-lines in close proximity to the MIT (where they will become particularly dense) and analyze their behavior towards ${T\!\rightarrow\!0}$.

\subsection{\label{Motivation} Limitations of existing results}
The occurrence of divergences of the irreducible vertex functions in several fundamental many-electron problems has been recently investigated in different publications \cite{Schaefer2016,Vucicevic2018,Chalupa2018,Thunstroem2018,Melnick2020,Springer2020}. In particular, one of the most systematic analysis made for the case of the half-filled HM on a square lattice solved in DMFT (essentially equivalent \cite{DOS_choice} to the case considered here) has been reported in \cite{Schaefer2016}, whose results are summarized in the central/right most panels of Fig.~\ref{fig:intro}. Specifically, the right panel of the figure, a zoom of the parameter region marked by a black box in the DMFT phase-diagram of the central panel, shows the first few $\boldsymbol{\Gamma}_c^\infty$-lines of the DMFT solution of the HM, marked in red or orange, depending whether the corresponding vertex divergence occurs in the charge sector only or, simultaneously, in the charge and in the particle-particle channels \cite{PP_divergencies}. 
As it was already noted \cite{Schaefer2016}, all the  $\boldsymbol{\Gamma}_c^\infty$-lines tend to approach the corresponding vertex divergences lines of the Hubbard atom (straight dashed lines starting from ${U\!=\!0}$) for large values of $T$ and $U$, while displaying a clear bending around the Mott MIT.

Evidently the clear bending of the first divergences lines, and their somewhat similar shape as the rightmost border of the MIT hysteresis $U_{c_2}(T)$, may suggest the originally proposed interpretation of the occurrence of the vertex divergences as {\sl precursors} of the Mott MIT itself.  
However, as already mentioned in the Introduction, subsequent studies have demonstrated the occurrence of similar vertex divergences in models, such as the AIM \cite{Chalupa2018}, where no Mott MIT takes place, and ascribed \cite{Gunnarsson2016,Chalupa2018,Chalupa2021,Adler2022} them to suppressive effects of the on-site/impurity charge fluctuation triggered by the formation of a local magnetic moment. 

Hence, in order to clarify the nature of the relation linking  the Mott MIT and the occurrence of divergences in the irreducible vertex functions of the charge sector, it is necessary to extend the DMFT studies performed hitherto to the most challenging parameter regime, namely, the coexistence region across the MIT, which represents the central goal of the present work.

\subsection{Methods}
The DMFT calculations of the generalized local charge susceptibility have been performed by using a continuous-time quantum Monte Carlo (CT-QMC) solver \cite{Gull2011} to sample the one and two-particle Green's functions in Eq.~(\ref{eq:chitau}) for the auxiliary AIM  of the corresponding self-consistent DMFT solutions. Specifically, we used the CT-QMC solver of the \textit{w2dynamics} package \cite{Wallerberger2019,Kowalski2019}. Further technical details about the numerical calculations are shortly reported in  Appendix~\ref{Workflow}. Here, we want to concisely recall, how the PM and PI DMFT-solutions in the coexistence region are obtained: Starting from  outside of the coexistence region, the interaction $U$ is changed step-by-step for a fixed temperature $T$, whereby the previously converged DMFT calculations are used as a starting point of the new self-consistent DMFT cycle (as schematically illustrated by the two arrows in the leftmost panel of Fig.~\ref{fig:intro}). By entering the coexistence/hysteresis region, the variation-steps in $U$ must be small (e.g. $\mathcal{O}(0.1)-\mathcal{O}(0.01)$), in order to allow for convergence to different meta-stable DMFT solutions, depending on the initial condition used. In this way two different solutions can be stabilized, at a given temperature $T$, in the interval ${U_{c1}(T)\!<\!U\!<\!U_{c2}(T)}$, the PM (PI) one being obtained along the path from the left to the right ${U_{c1}(T)\!\to\!U_{c2}(T)}$ (the right to the left ${U_{c2}(T)\!\to\!U_{c1}(T)}$). For each PM or PI converged DMFT-solution obtained in the coexistence region, the corresponding on-site generalized charge susceptibility is then computed as explained above, and Fourier transformed in Matsubara frequencies. The diagonalization of their corresponding matrix representation in the fermionic Matsubara frequencies allows to determine  the number of negative eigenvalues, $N_{\lambda<0}$, which, as discussed in Sec.~II B, corresponds to the number of crossed $\boldsymbol{\Gamma}_c^\infty$-lines and can then be used to approximate the $\boldsymbol{\Gamma}_c^\infty$-lines (see Appendix \ref{Approx_of_lines} for further details) in the region of the phase-diagram close to the Mott MIT.

\section{\label{Results}Results}
\subsection{\label{Metallic_results}Metallic coexistence region}

\begin{figure}[b]
\includegraphics[width=\linewidth]{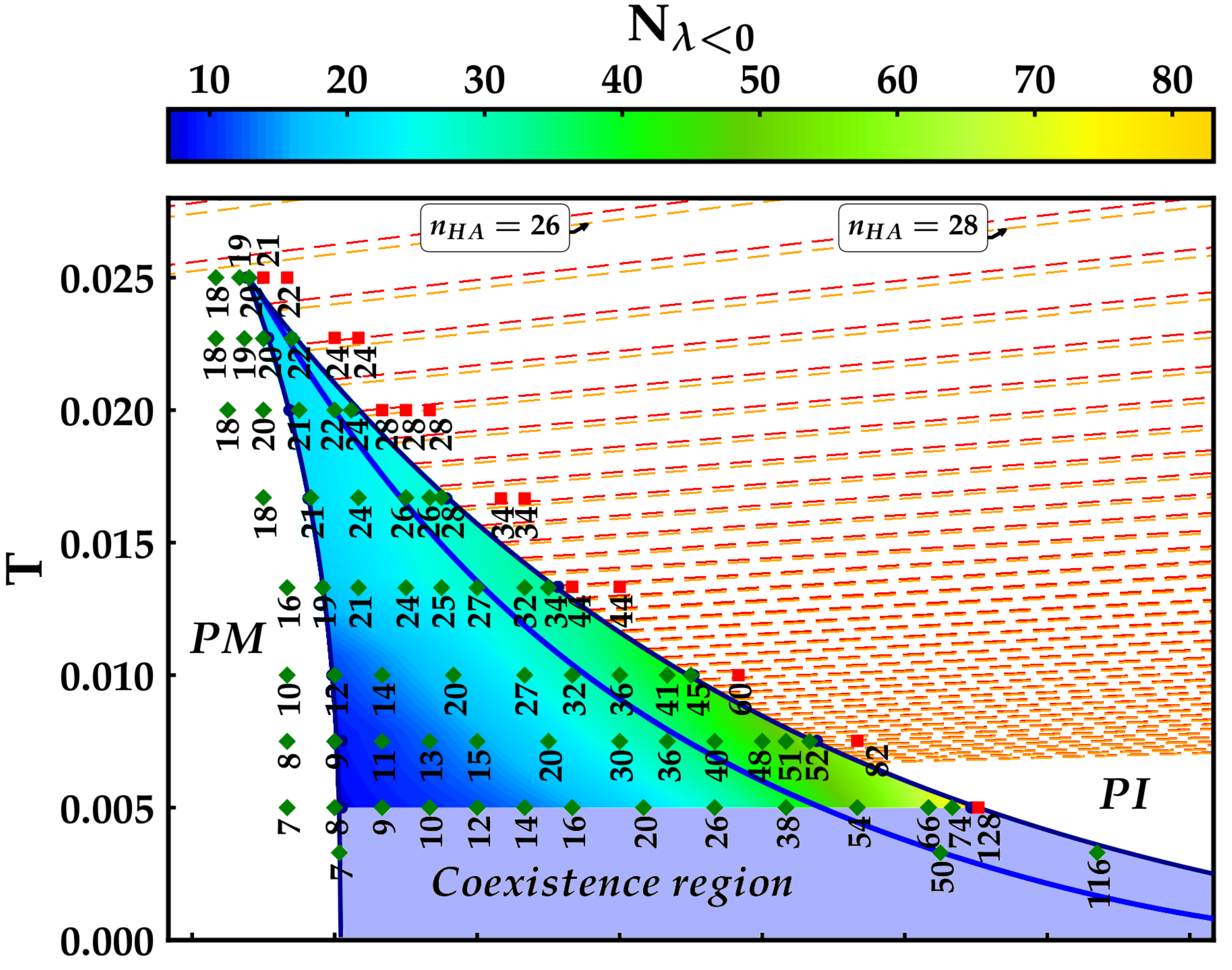}
\includegraphics[width=\linewidth]{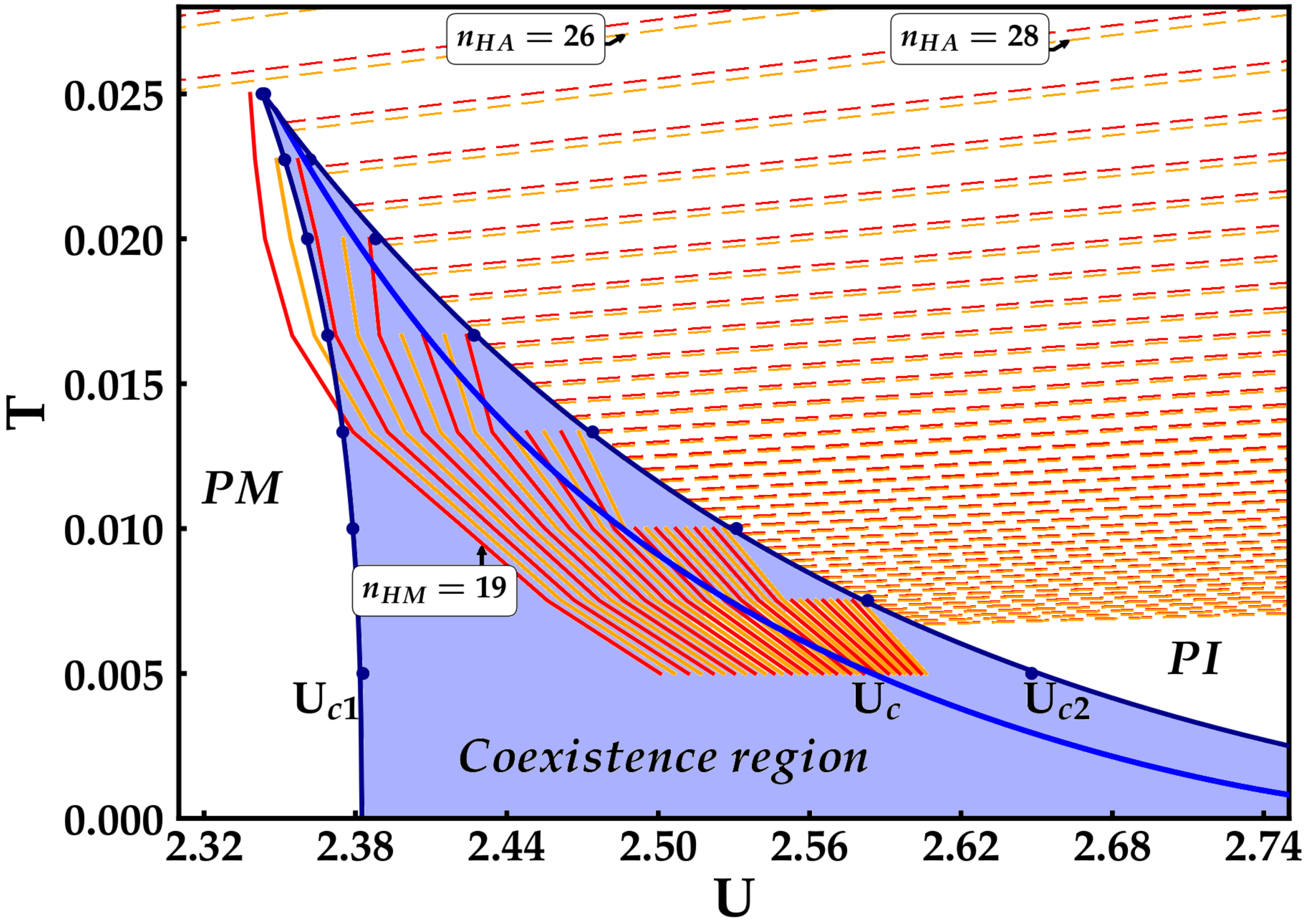}
\caption{\label{fig:MetallicCoex}Phase diagrams of the MIT with PM solution in the coexistence region (blue-shaded area) for the Hubbard model (HM) on the Bethe lattice. $U_c$ (blue), taken from Ref.~\cite{Blumer2002}, denotes the thermodynamic transition. Upper panel: Coexistence region with phase points of performed DMFT calculations, where green diamonds correspond to a metallic solution and red squares to an insulating one. The numbers next to markers are $N_{\lambda<0}$ and the background of the points within the coexistence region shows an interpolating color scale of $N_{\lambda<0}$ for the metallic solution. Lower panel: Same phase diagram as the upper panel, but showing the distinct $\boldsymbol{\Gamma}_c^\infty$-lines approximated from the data of the phase points (see Appendix \ref{Approx_of_lines}). Here, $n_{HM}$ indicates the number of crossed $\boldsymbol{\Gamma}_c^\infty$-lines of the Hubbard model, coming from ${U\!=\!0}$. Dashed red and orange lines (in both panels) mark the $\boldsymbol{\Gamma}_c^\infty$-lines of the Hubbard atom (HA) according to \cite{Schaefer2016} as reference, where $n_{HA}$ is the number of crossed lines coming from ${U\!=\!0}$.}
\end{figure}

Our results for the PM solutions of the MIT coexistence region are shown in the upper panel of Fig.~\ref{fig:MetallicCoex}. Here the coexistence region is indicated as a blue framed and shaded area in the $T$-$U$ phase diagram. The two-particle calculations performed by scanning the MIT starting from the PM side are marked, respectively, with green diamonds or red squares, depending on whether a PM or a PI solution is found (the stabilization of the first PI solutions evidently corresponds to the crossing of the $U_{c2}(T)$ line). The calculated number $N_{\lambda<0}$ of negative eigenvalues of $\boldsymbol{\chi}_c^{\nu\nu^\prime}$ is shown right below the corresponding markers. In the background of the data points, a color scale, interpolating between the numerical results, indicates the change of $N_{\lambda<0}$ within the coexistence region. We emphasize here that, in order to make the comparison between the results obtained in the different parameter regimes more easily accessible, the {\sl same} color scale has been used for all intensity-plots shown in the paper.

In the lower panel of Fig.~\ref{fig:MetallicCoex} instead, we show the corresponding vertex divergence  ($\boldsymbol{\Gamma}_c^\infty$) lines, starting with the $\boldsymbol{\Gamma}_c^\infty$-line associated to ${N_{\lambda<0}\!=\!n_{HM}\!=\!19}$, and we adopt the same (red and orange) color-coding \cite{color_coding} as introduced in Sec.~II C. As reference for the Mott insulating phase, the $\boldsymbol{\Gamma}_c^\infty$-lines of the Hubbard atom (HA) \cite{Schaefer2016,Thunstroem2018} are shown as dashed red and orange lines on the right (PI) side of the coexistence region, starting with ${N_{\lambda<0}\!=\!n_{HA}\!=\!26}$.

In both panels of Fig.~\ref{fig:MetallicCoex} a clear bending of the $\boldsymbol{\Gamma}_c^\infty$-lines towards the critical point of the Mott MIT at ${T\!=\!0}$ ($U_{c2}^{T=0}$) is observed in the whole PM coexistence region, corresponding to an increase of $N_{\lambda<0}$ along an increase of $U$ and $T$. One also notices that the $\boldsymbol{\Gamma}_c^\infty$-lines occur particularly dense for increasing $U$ at low temperatures. More specifically, the shape of the bending reminds on the first $\boldsymbol{\Gamma}_c^\infty$-lines encountered
in the correlated metallic regime of the DMFT solution of the HM \cite{Schaefer2016}
(central panel in Fig.~\ref{fig:intro}), as well as the $\boldsymbol{\Gamma}_c^\infty$-lines found in low-$T$ region of the phase-diagram of the AIM \cite{Chalupa2018}. On a more physical perspective, the $\boldsymbol{\Gamma}_c^\infty$-lines in the PM coexistence region display a qualitative similarity with the Mott transition line $U_c$ (solid blue line in Fig.~\ref{fig:MetallicCoex}), especially for low temperatures,
while no significant match with the effective \cite{Mazitov2022} Kondo temperature associated to the auxiliary AIM of the DMFT solution (estimated through the so-called ``fingerprints criterion" \cite{Chalupa2021}) can be noticed. These observations suggest the existence of a direct connection between the vertex divergences and the Mott MIT itself.

\subsection{\label{Insulating_results}Insulating coexistence region}

\begin{figure}[t]
\includegraphics[width=\linewidth]{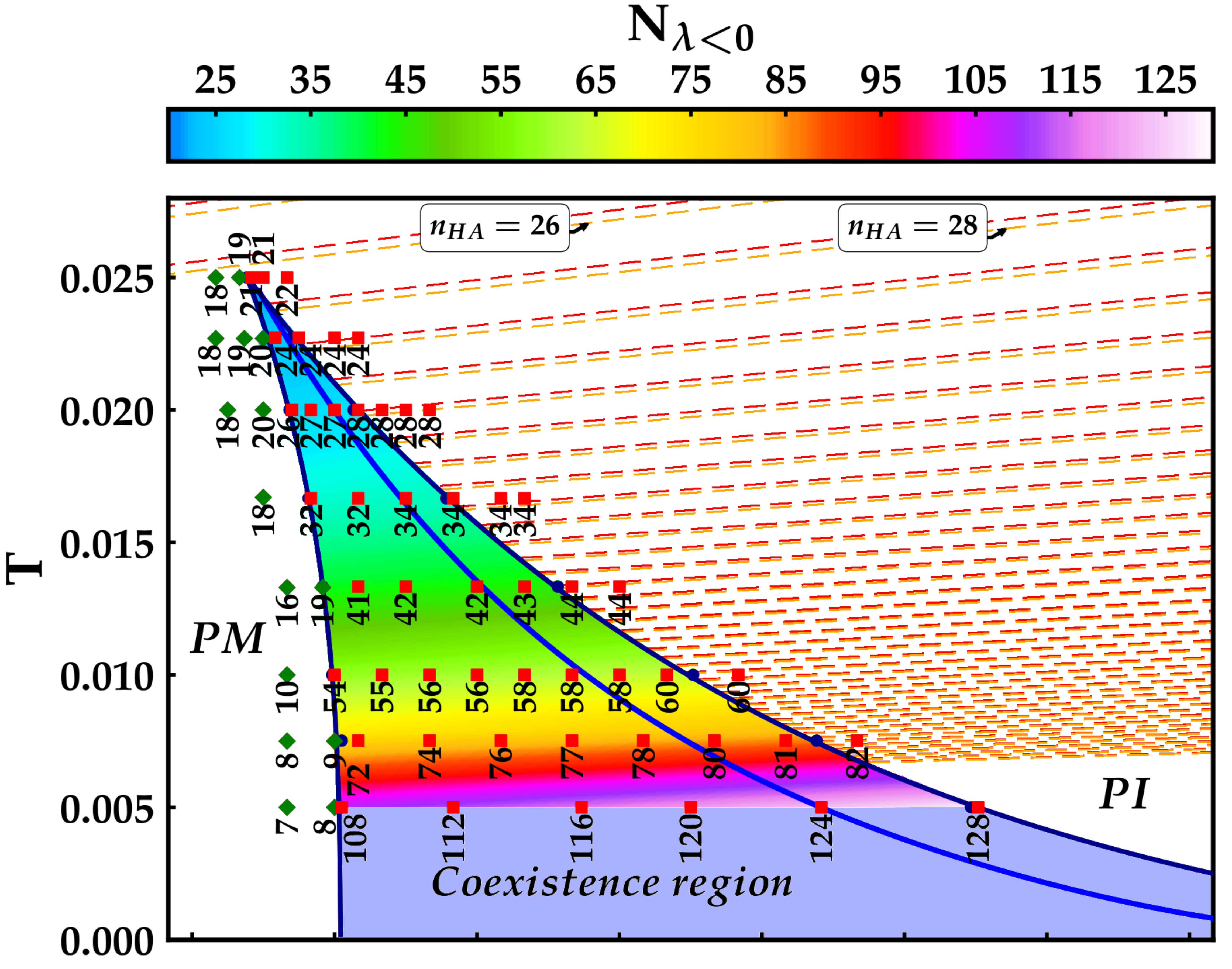}
\includegraphics[width=\linewidth]{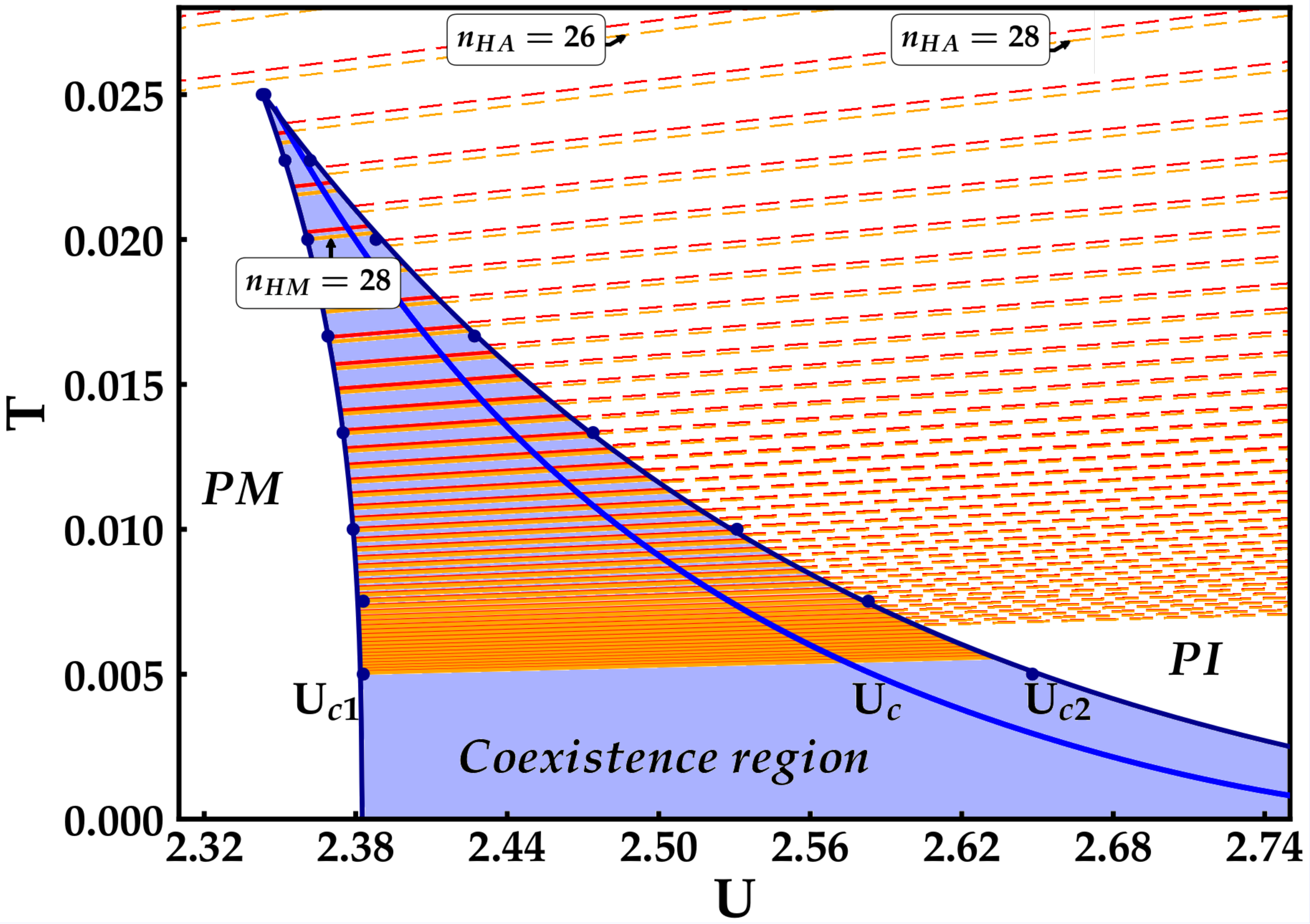}
\caption{\label{fig:InsulatingCoex}Phase diagrams of the MIT with PI solution in the coexistence region (blue-shaded area) for the Hubbard model (HM) on the Bethe lattice. The blue line $U_c$, taken from Ref.~\cite{Blumer2002}, denotes the thermodynamic transition. Upper panel: The data points, red squares for an insulating phase and green diamonds for a metallic phase, mark the performed DMFT calculations. The number next to the points shows the corresponding $N_{\lambda<0}$ and the background of the data points in the coexistence region shows an interpolating color scale of $N_{\lambda<0}$. Lower panel: Same phase diagram as the upper panel with approximated $\boldsymbol{\Gamma}_c^\infty$-lines of the HM, $n_{HM}$ is the number of crossed lines coming from ${U\!=\!0}$. Dashed red and orange lines (in both panels) mark the $\boldsymbol{\Gamma}_c^\infty$-lines of the Hubbard atom (HA) according to \cite{Schaefer2016} as reference, where $n_{HA}$ is the number of  $\boldsymbol{\Gamma}_c^\infty$-lines of the HA.}
\end{figure}

Our results for the PI solutions of the MIT coexistence region are shown in the upper panel of Fig.~\ref{fig:InsulatingCoex}. The two-particle calculations performed by scanning the MIT starting from the PI side are marked, respectively, with red squares or green diamonds, depending on whether a PI or a PM solution is found (the stabilization of the first PM solutions evidently corresponds to the crossing of the $U_{c1}(T)$ line).

\begin{figure*}[t]
\includegraphics[width=\linewidth]{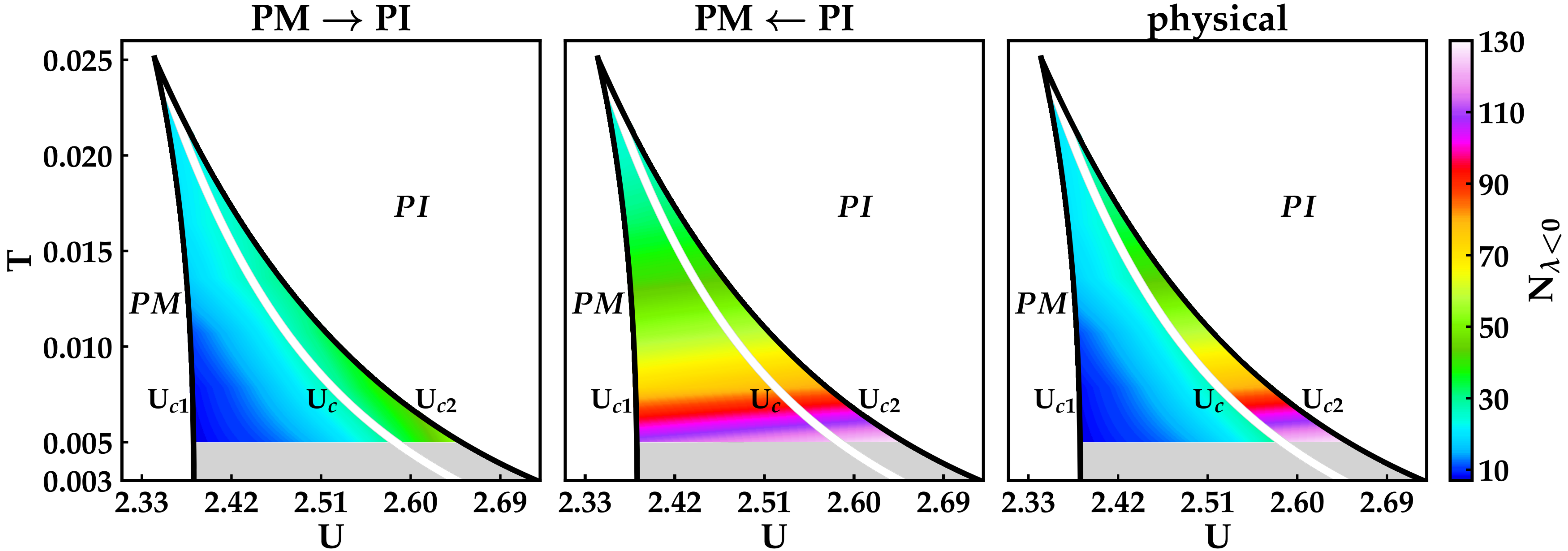}
\caption{\label{fig:met-ins-phys}$T$-$U$ diagram of the MIT coexistence regions for the Hubbard model on the Bethe lattice showing the number of negative eigenvalues of the generalized charge susceptibility $N_{\lambda<0}$ as interpolating color scale for the metallic solution (left), the insulating solution (middle) and comparing $N_{\lambda<0}$ at the thermodynamic stable (physical) transition $U_c$ of Ref.~\cite{Blumer2002} (right panel).}
\end{figure*}

As before, the color scale shows the interpolation of $N_{\lambda<0}$ in the coexistence region. In the lower panel of Fig.~\ref{fig:InsulatingCoex}, the distinct $\boldsymbol{\Gamma}_c^\infty$-lines (orange/red) are shown. The $\boldsymbol{\Gamma}_c^\infty$-lines are now {\sl straight} lines, similar to the dashed (orange/red) lines of the HA (i.e. the extreme case of an HM with ${t\!=\!0}$),   shown as reference on the right side of the coexistence region. Due to the different shape of the divergences lines w.r.t.~the PM solution analyzed before, $N_{\lambda<0}$ still increases with increasing $U$, but, differently from the PM case of Fig.~\ref{fig:MetallicCoex}, it increases for decreasing $T$.

At the same time, the qualitative similarity between the 
$\boldsymbol{\Gamma}_c^\infty$-line shape in the HA and the PI phase of the HM solved in DMFT cannot represent, evidently, an identity \cite{DelRe2021}, reflecting the intrinsic difference between the ``perfect" localization of the HA and the actual one of the Mott PI phase, where finite (though quite small) double-occupancy is found even in the ground-state.
In fact, by comparing the results for the HA and the PI phase in DMFT, on a quantitative level, (e.g. by considering the 28-th line of the HA ${n_{HA}\!=\!28}$, and the corresponding line of the HM ${N_{\lambda<0}\!=\!n_{HM}\!=\!28}$ in the lower panel of Fig.~\ref{fig:InsulatingCoex}) a systematic shift of HM-$\boldsymbol{\Gamma}_c^\infty$-lines  w.r.t.~the HA-$\boldsymbol{\Gamma}_c^\infty$-lines can be noted. To effectively account for this shift, one might rescale the interaction of the HA \cite{HA_rescaling}, by comparing the number of $n_{HA}$ to $n_{HM}$ for all phase points within the coexistence region, which yields an average factor of ${\eta\!=\!\frac{n_{HA}}{n_{HM}}\!=\!1.19\pm0.03}$ (see Appendix \ref{effectiveHA} for further details). Hence, at least in terms of vertex divergences, it might be tempting to ``approximate" the PI phase of the HM in a similar spirit as in Ref.~\cite{DelRe2021},  as HA with ``reduced" effective interaction $U_{eff}$:
\begin{equation}
    \mathcal{H}_{PI} = U_{eff}\sum_{i}n_{i\uparrow}n_{i\downarrow} \text{\ \ with\ \ } U_{eff}=\frac{U}{\eta}.
	\label{HMPI}
\end{equation}
Obviously, this analogy cannot be pushed too far: In the limit of ${U\!\rightarrow\!\infty}$,  the HM-$\boldsymbol{\Gamma}_c^\infty$-lines  will gradually approach the HA-$\boldsymbol{\Gamma}_c^\infty$-lines asymptotically, i.e., ${U_{eff}\!\rightarrow\!U}$ and ${\eta\!\rightarrow\!1}$ for $U\!\rightarrow\!\infty$. In practice, ${\eta \approx\!1.19}$ can be used as a good approximation for the vertex divergences of the PI phase of the HM, in the (physically relevant) regime of the MIT \cite{MIT_eta}.

\subsection{\label{VertDivTrans} Vertex divergence lines at the phase transition}

After separately analyzing how vertex divergences occur in the two (PM and PI) sets of DMFT solutions in the coexistence region, the question of their behavior across the thermodynamic transition can be now readily addressed.
We briefly recall that the actual thermodynamic phase transition takes place at $U\!=\!U_c(T)$ (cf.~Ref.~\cite{Blumer2002}), when the free energy of the PM and the PI converged DMFT-solutions become equal.
The corresponding transition line \cite{Blumer2002} marked as white line in all three panels of Fig.~\ref{fig:met-ins-phys}, thus separates the thermodynamically {\sl stable} PI and PM solutions (labeled as physical in Fig.~\ref{fig:met-ins-phys}) on the two sides of the coexistence region and is associated, except at its second-order (quantum) critical endpoints, to an abrupt (first-order) jump of the physical properties from a PM to a PI behavior, except at its second-order (quantum) critical endpoints.

After these premises, we summarize in Fig.~\ref{fig:met-ins-phys} our results for the vertex divergences in the proximity of the Mott MIT. In the first two panels, the respective values $N_{\lambda<0}$ for the PM (left panel) and the PI (middle panel) solutions are shown as color-intensity plots, making the qualitatively different shape of the $\boldsymbol{\Gamma}_c^\infty$-lines between the PM and the PI quite evident. From the direct comparison of the first two panels, the quantitative difference in $N_{\lambda<0}$ becomes also clearly visible, as the values of $N_{\lambda<0}$ appear to be systematically higher in the PI than in the PM realization of the coexistence region.

\begin{figure*}[t]
\includegraphics[width=\linewidth]{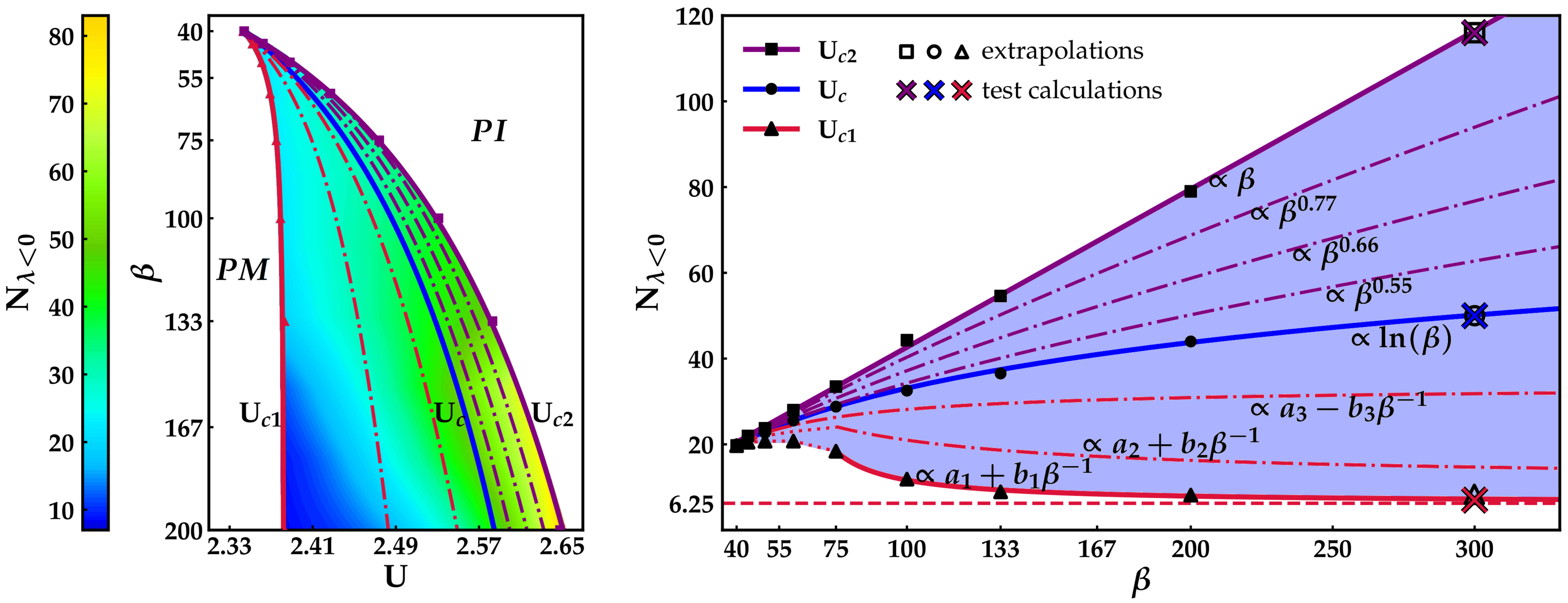}
\caption{\label{fig:Nvert_beta_met}Left panel: Phase diagram of the inverse temperature ${\beta\!=\!(k_\mathrm{B}T)^{-1}}$ and interaction $U$ showing the MIT and the coexistence region with an interpolating color scale of the number of negative eigenvalues of the generalized charge susceptibility $N_{\lambda<0}$ for the metallic phase, where the thermodynamic transition line $U_c$ is taken from Ref.~\cite{Blumer2002}. Right panel: $N_{\lambda<0}(\beta)$; black markers are calculated values of $N_{\lambda<0}$ for $U_{c1}$ (triangles), $U_{c}$ (circles) and $U_{c2}$ (squares) according to the left panel. The solid lines correspond to fits for these data points, yielding ${N_{\lambda<0}^{U_{c1}}(\beta)\!\cong\!a_1\!+\!b_1\beta^{-1}}$ \cite{fitting_function} for $\beta\rightarrow\infty$ with the red dashed line as asymptotic value (the red dotted lines are a guide to the eye), ${N_{\lambda<0}^{U_{c}}(\beta)\!\propto\!\ln(\beta)}$ and ${N_{\lambda<0}^{U_{c2}}(\beta)\!\propto\!\beta}$. Black empty markers are estimated values from the fits for ${\beta\!=\!300}$ and colored crosses are the results of corresponding DMFT test calculations. The dashed dotted lines are intermediate lines ($U^{int}_i$) interpolating between $U_{c1}$, $U_{c}$, and $U_{c2}$.}
\end{figure*}

Hence, when directly considering the vertex divergence behavior for the thermodynamic transition, as we do in the rightmost panel of Fig.~\ref{fig:met-ins-phys}, $N_{\lambda<0}$ displays an evident jump  $\Delta N_{\lambda<0}$ between the PM and PI, at the first-order transition $U_c(T)$. On a more quantitative level, we note that $\Delta N_{\lambda<0}$ between the PM and the PI along $U_c$ increases with decreasing temperature. Only at the second order critical point at ${T\!=\!T_c\!\approx\!\frac{1}{39}}$ we observe a continuous transition of $N_{\lambda<0}$ with ${\Delta N_{\lambda<0}\!=\!0}$, as expected. In this regard, $\Delta N_{\lambda<0}$ appears to reflect well the behavior of the order parameter characterizing the transition.
As discussed in the next subsection, however, the same analogy will not directly apply to the other extreme ($T=0$) of the Mott MIT line, as this point will be characterized by a {\sl divergence} of $N_{\lambda<0}$.

\subsection{\label{Accumulation}Accumulation of vertex divergence lines}

Our calculations for $N_{\lambda<0}$ in the coexistence region at finite temperatures naturally raise the question of how $N_{\lambda<0}$ will behave at temperatures lower than those accessible by our numerical calculations and, in particular, for ${T\!\rightarrow\!0}$. To this aim, we performed numerical extrapolations for $N_{\lambda<0}(T)$ from our data and tested the validity of those by comparing them to additional, numerically heavier calculations, performed at a ``testbed'' temperature value (e.g., $\beta =300$ for the PM-phase) lower than the temperature-interval used for the extrapolation. 

\begin{figure*}[t]
\includegraphics[width=\linewidth]{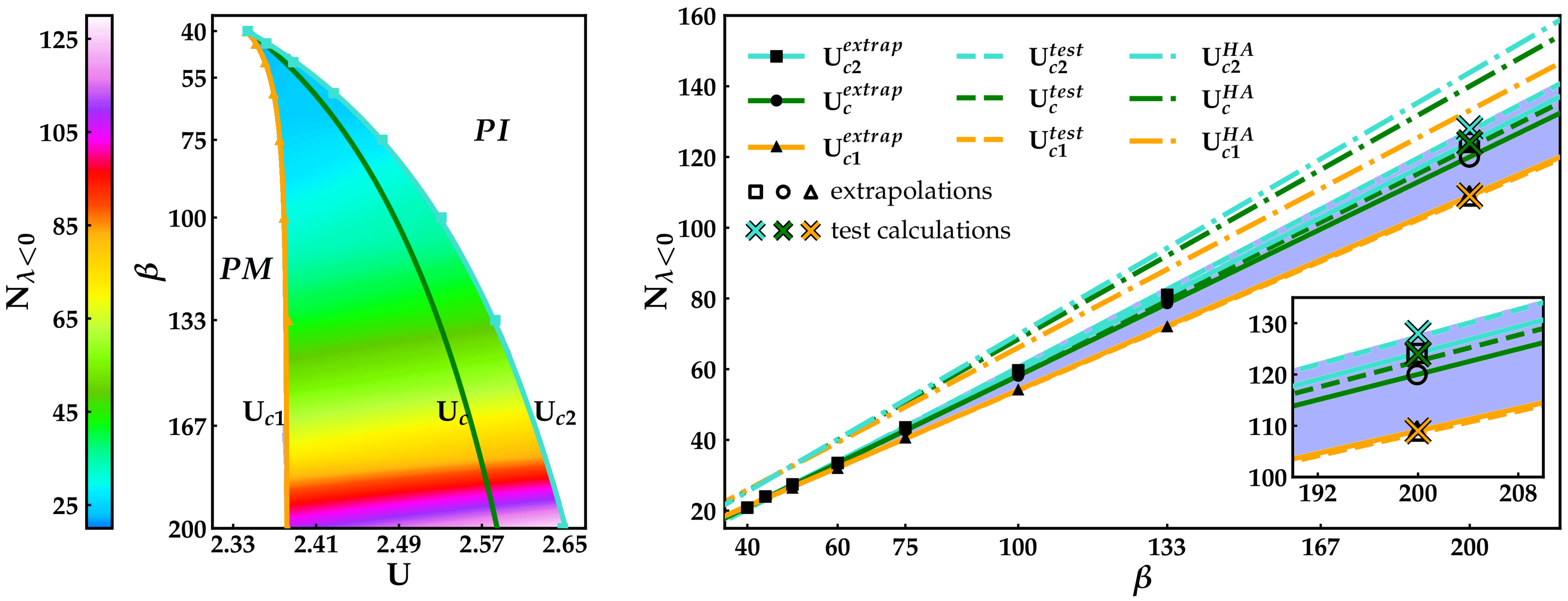}
\caption{\label{fig:Nvert_beta_ins}Left panel: Phase diagram of the MIT and its coexistence region with an interpolating color scale of the number of negative eigenvalues $N_{\lambda<0}$ of the generalized charge susceptibility for the insulating results. The thermodynamic transition line $U_c$ is taken from \cite{Blumer2002}. Right panel: $N_{\lambda<0}(\beta)$; black markers are determined values of $N_{\lambda<0}$ for $U_{c1}$ (triangles), $U_{c}$ (circles) and $U_{c2}$ (squares) according to the left panel. The solid lines mark fits of these data, yielding linear behavior for the three transition lines. Black empty markers are estimated values from these fits for ${\beta\!=\!200}$ and colored crosses are the results of corresponding test calculations (a close-up is shown in the inset on the right). The dashed lines are fits including the test calculations and the dashed dotted ones present the number of crossed $\boldsymbol{\Gamma}_c^\infty$-lines of the Hubbard atom along $U_{c1}$, $U_{c}$ and $U_{c2}$}
\end{figure*}

Our results for the PM solution are shown in  Fig.~\ref{fig:Nvert_beta_met}. Specifically, in the left panel, we show, as a guidance, $\beta$-$U$ paths in the coexistence region, along which our extrapolation are performed, superimposed to the corresponding color-intensity plot for (the numerically interpolated) $N_{\lambda<0}$. In the right panel, then, we report our numerically calculated data $N_{\lambda<0}$ as function of the inverse temperature $\beta$ (black markers) together with the corresponding fits along $U_{c1}$, $U_{c}$, $U_{c2}$ (solid lines) as well as along intermediate (dash dotted) paths in parameter space (see Appendix \ref{intermediate_lines} for further details). The range of the $N_{\lambda<0}(\beta)$ values in the coexistence region is indicated as blue-shaded area. As mentioned before, the reliability of our extrapolation has been tested by comparing additional calculations at ${\beta\!=\!300}$ (crosses) to the extrapolated values (empty markers), which yielded a good agreement.

By a closer inspection of the results, we note the following differences in the ${T\!\rightarrow\!0}$ behavior of $N_{\lambda<0}$ along the distinct paths we selected. For instance, starting with the transition line $U_{c1}(\beta)$ we find ${N_{\lambda<0}\!\cong\!a_1\!+\!b_1\beta^{-1}}$ \cite{fitting_function} for ${\beta\!\rightarrow\!\infty}$ (${a_1\!\approx\!6.25}$), which would be compatible with ${N_{\lambda<0}\!=\!6}$ at $U_{c1}(T\!=\!0)$. Along the intermediate lines at ${U_{c1}\!<\!U\!<\!U_c}$ we find ${N_{\lambda<0}\!\cong\!a_i\!+\!b_i\beta^{-1}}$ with finite ${a_i\!>\!a_1}$. At the transition lines $U_c(\beta)$ and $U_{c2}(\beta)$, which both end at the second-order critical point at $U_{c2}(T=0)$, $N_{\lambda<0}(\beta)$ is observed to grow {\sl logarithmically} and {\sl linearly} in $\beta$, respectively. Consistently, the intermediate lines in the interval ${U_c\!<\!U\!<\!U_{c2}}$ feature a divergent behavior of ${N_{\lambda<0}(\beta)\!\propto\!\beta^x}$ with ${0\!<\!x\!<\!1}$. This indicates that an {\sl infinite} number of $\boldsymbol{\Gamma}_c^\infty$-lines need to be crossed before reaching $U_{c2}(T\!=\!0)$, i.e., the location of the Mott-Hubbard MIT in the ground-state.

Our extrapolations for $N_{\lambda<0}(\beta)$ of the PI solution are displayed Fig.~\ref{fig:Nvert_beta_ins}. Again, in the left panel of the figure the respective location of the paths considered in the $\beta$-$U$ coexistence region are shown, superimposed to the corresponding values of $N_{\lambda<0}(\beta)$ for the PI solutions. Due to the numerical hurdles of performing two-particle DMFT calculations for low $T$ in the PI-phase (see Appendix \ref{sampling}), the extrapolations (solid lines) and the test calculations (empty markers) have been performed at higher temperatures than for the PM phase (PI test calculations at ${\beta\!=\!200}$), which likely explains the overall less satisfactory agreement between extrapolation and test calculations w.r.t.~the PM case. The inset in the right panel presents a close-up of the corresponding deviations. Fits including the $\beta=200$ calculations, which should be regarded as the most precise results which we could obtain in the PI phase, are displayed as dashed lines. As comparison, dashed dotted lines present $N_{\lambda<0}(\beta)$ of the HA along $U_{c1}$, $U_{c}$, and $U_{c2}$, all showing higher values than the PI solution of the HM in the coexistence region. 

The extrapolations for $N_{\lambda<0}(\beta)$ in the PI (with and without test calculations) display a linear behavior in $\beta$ for all three transition lines $U_{c1}$, $U_{c}$, $U_{c2}$, corresponding to an infinite number of $\boldsymbol{\Gamma}_c^\infty$-lines to be crossed before reaching ${T\!=\!0}$. On a more quantitative level, we note that a comparison between the slopes $\alpha_i$ of the extrapolations of $N_{\lambda<0}(\beta)$ and the corresponding results of the HA, can be also used to obtain ${\eta\!=\!\frac{\alpha_{HA}}{\alpha_{HM}}}$ for an effective HA like description of the PI solution of the HM in Eq.~(\ref{HMPI}) of Sec.~III~B, e.g.~${\eta\!=\!1.22219}$ for $U_{c1}(T)$ (for details see Appendix \ref{effectiveHA}).

From our extrapolations of $N_{\lambda<0}(\beta)$ we can now try to estimate the overall behavior of $N_{\lambda<0}$ at ${T\!=\!0}$ across the MIT. By crossing the MIT from ${\text{PM}\!\rightarrow\!\text{PI}}$ at ${T\!=\!0}$, the number of $\boldsymbol{\Gamma}_c^\infty$-lines increases to infinity at the critical endpoint $U_{c2}(T\!=\!0)$. In this respect, at ${T\!>\!0}$ the physical transition line $U_c(T)$ can bee seen as the first of the intermediate line-paths in $\beta$,$U$ of Fig.~\ref{fig:Nvert_beta_met} along which $N_{\lambda<0}(\beta)$ diverges for ${\beta\!\rightarrow\!\infty}$, namely logarithmically. In the metastable metallic phase for ${U_c\!<\!U\!<\!U_{c2}}$ -- along the intermediate lines -- $N_{\lambda<0}(\beta)$ can diverge faster than $\log{\beta}$, and finally ${N_{\lambda<0}(\beta)\!\propto\!\beta}$ at $U_{c2}$. The same $\beta$-dependence, namely ${N_{\lambda<0}(\beta)\!\propto\!\beta}$, occurs in the PI coexistence region as well as in the HA. 

On the basis of this numerical evidence, we can then conclude that at the Mott-Hubbard MIT at ${T\!=\!0}$ an accumulation point of an {\sl infinite} number of vertex divergence lines occurs, a clear link between the Mott MIT at ${T\!=\!0}$ and the vertex divergence lines.
Clearly, the different paths in the $\{U, \beta \}$ parameter plane, through which the $T\!= \! 0$ transition can be approached,  e.g.~ $U_{c}(\beta)$ or $U_{c2}(\beta)$, reflect in correspondingly different ways in which the {\sl divergence} of $N_{\lambda<0}(\beta \rightarrow \infty)$ is eventually achieved \footnote{Evidently, this also means that the difference of $N_{\lambda<0}(\beta)$  between  different paths approaching the $T\!=\!0$ MIT, e.g. $U_{c}(\beta)$ or $U_{c2}(\beta)$,  does not necessarily vanish (and it might even diverge) for $\beta \rightarrow \infty$, depending on the pair of paths chosen.}. However, as the asymptotic value of $N_{\lambda<0}(\beta \rightarrow \infty)$ always diverges when approaching $U_c(T\!=\!0) \!= \! U_{c2}(T\!=\!0)$ independently of the path, our extrapolated results for $N_{\lambda<0}(\beta \rightarrow \infty)$ appear consistent \footnote{In particular, the $T\!=\!0$-extrapolated behavior of $N_{\lambda<0}$ could be consistent with the existence of a unique function describing the increase of $N_{\lambda<0}$ for increasing $U$ along the $T\!=\!0$-axis up to its divergence at ${U=U_{c2}(T\!=\!0)}$, and, hence, does not rule out the possibility of a smooth evolution of the generalized susceptibility  $\boldsymbol{\chi}^{\nu  \nu'}_c$ and of the corresponding physical response function.}  with the smooth evolution of several physical quantities (such as the quasiparticle spectral weight, the double-occupancy, etc.) numerically reported \cite{Karski2005, Raas2009} at the $T=0$-Mott MIT.

\section{Theoretical and Algorithmic implications}

\begin{figure*}[t]
\includegraphics[width=\linewidth]{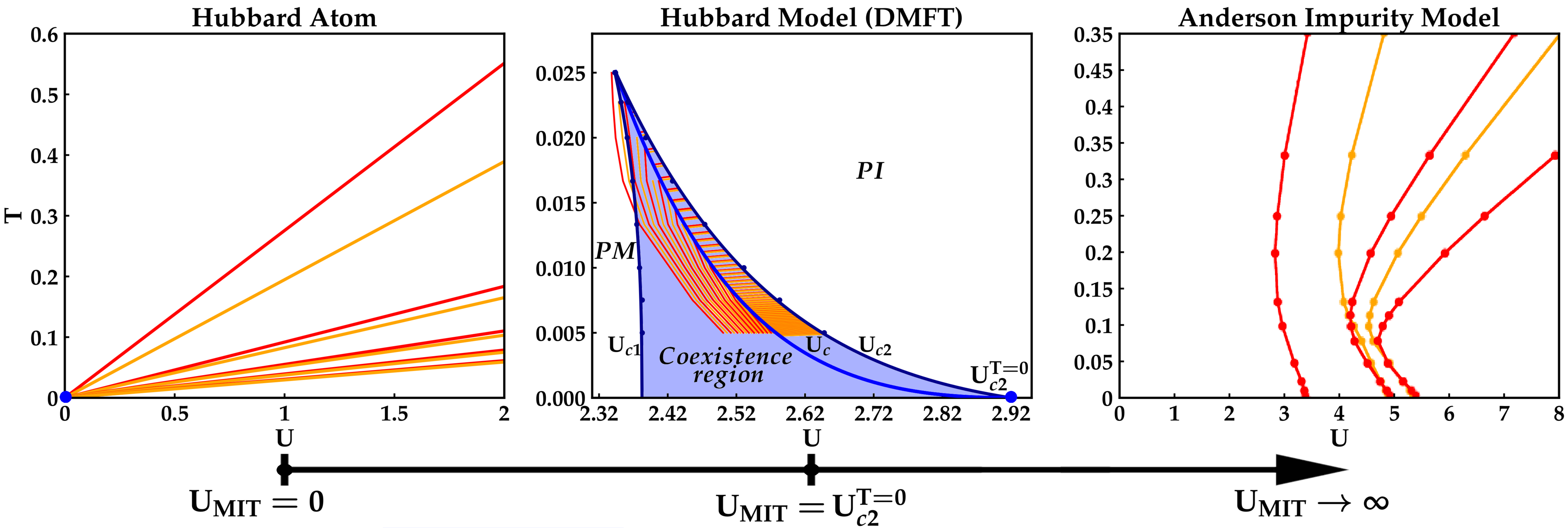}
\caption{\label{AL_HM_AIM} Phase diagrams of the Hubbard atom (HA) (reconstructed from Ref.~\cite{Schaefer2016}), Hubbard model (HM) and Anderson impurity model (AIM) (reconstructed from Ref.~\cite{Chalupa2018}) with the corresponding $\boldsymbol{\Gamma}_{c}^\infty$-lines and their accumulation point (blue dot) at ${T\!=\!0}$ (${U_{MIT}\!\!0}$ for the HA, ${U_{MIT}\!=\!U_{c2}^{T=0}}$ for the HM and ${U_{MIT}\!\rightarrow\!\infty}$ for the AIM). The black arrow below sketches the shift of the accumulation point of the $\boldsymbol{\Gamma}_{c}^\infty$-lines between the different models. The accumulation point at the MIT of the DMFT solution of the Hubbard model at $U_{c2}(T\!=\!0)$ can be seen as intermediate case w.r.t.~the \textit{``extreme"} cases of the purely insulating Hubbard atom and the purely metallic Anderson impurity model.}
\end{figure*}

Our analysis of the irreducible vertex divergences in the close proximity of the Mott MIT has several important implications both of algorithmic as well as of conceptual nature. We demonstrated the existence of a large number of divergence lines found in the coexistence region of the MIT, increasing up to infinity along all calculated parameter paths towards the ${T\!=\!0}$ MIT at ${U_{c2}(T\!=\!0)}$ evidently poses a huge  challenge to the applicability of all algorithmic approaches using and/or explicitly manipulating irreducible vertices of DMFT as input, such as the full-fledged, parquet-equation based version of the dynamical vertex approximation \cite{Toschi2007,Valli2015,Rohringer2018,Kauch2020,Kaufmann2021} and the QUADRILEX scheme \cite{Ayral2016}.
In fact, while one could realistically cope with a situation of a few well-separated divergence lines, it becomes hard to stabilize the numerical manipulation of irreducible vertices in parameter regimes, where their frequency structure will display large oscillations, due to the proximity of several and closely spaced divergence lines. 

At the same time, we recall that the occurrence of vertex divergences has been recently proven to be pivotal \cite{Chalupa2021,Adler2022} to ensure the correct transfer of physical information among the spin/magnetic and the charge/density or the pairing channels in the nonperturbative regimes, e.g., to feature the proper freezing of the local charge (or pairing) fluctuations associated to the pre-formation of local magnetic moments. Hence, it becomes clear why self-consistent diagrammatic approaches, whose irreducible vertex functions cannot diverge per construction \cite{simultaneous_divergencies}, such as the truncated functional renormalization group (fRG) \cite{Metzner2012} or the parquet approximation \cite{Bickersbook2004}, will not be able to yield a consistent picture of the Mott MIT and its related phenomena.

While it is beyond the scope of this work to address the issue of possible strategies how to include this relevant piece of physical information in diagrammatic treatments of the many-electron problems in nonperturbative regimes, e.g, by means of the merger of DMFT and fRG \cite{Taranto2014,Wentzell2015,Vilardi2019,Bonetti2020,Bonetti2022,Chalupa-Gantner2022} or by rewriting ladder diagrammatic resummations or parquet equations in terms of the full two-particle scattering amplitude of the local problem \cite{Hafermann2009,Rohringer2012,Rohringer2018,Krien2020b}, here we want to discuss some more fundamental theoretical implication of our results.

Indeed, the identification of the Mott-Hubbard MIT at zero temperature with the accumulation point of an {\sl infinite} number of $\boldsymbol{\Gamma}_c^\infty$-lines does not only clarify the precise relation linking the occurrence of vertex divergences in the charge channel and the MIT, highlighting the profoundly nonperturabtive nature of the MIT itself, but also allows to draw interesting, more general considerations about how the vertex divergences affect different many-electron problems.
In particular, our results for the HM in DMFT (summarized in the central panel of Fig.~\ref{AL_HM_AIM}) can be put into a broader perspective by directly comparing them to the corresponding ones for the Hubbard atom \cite{Schaefer2016,Thunstroem2018} (HA, left panel) and the (metallic) Anderson impurity model of Ref.~\cite{Chalupa2018} (AIM, right panel) in Fig.~\ref{AL_HM_AIM}, where we used the same color-coding for denoting  the different vertex divergence lines.

Thereby, we note that for the HA, i.e.~for an Hubbard model with ${t\!\equiv\!0}$, which features a perfect local-moment/insulating behavior for all ${U\!>\!U_{MIT} \!=\!0}$, an accumulation point of vertex divergence lines  occurs  exactly at the origin of the phase diagram, namely at ${U\!=\!T\!=\!0}$. On the other hand, by looking at the phase-diagram of the AIM, where no transition to an insulating ground state occurs for any value of $U$, we do not observe an accumulation point of vertex divergence lines in the available data \cite{Chalupa2018} and, in general, we do not expect one at ${T\!=\!0}$. We can note, nonetheless, that their low-$T$ distribution becomes denser by increasing interaction values.
One can then assume the asymptotic value ${U_{MIT}\!=\!+\!\infty}$ as an hypothetical value for the transition to an insulating state in the AIM. In this regime, the physics would be eventually dominated by $U$, similarly to the insulating phases of the other two cases, and the progressively denser distribution of vertex divergence lines by increasing $U$ could then lead asymptotically to an accumulation point.

Within this framework \cite{framework}, the results we obtained for the HM in DMFT can be naturally interpreted as an ``intermediate" case between the two extreme situations of the HA (with accumulation point at ${U_{MIT}\! =\!0}$) and the AIM (at ${U_{MIT}\!=\!+\!\infty}$), where $U_{MIT}$ yields the finite value of $U_{c2}(T\!=\!0)$, as schematically illustrated by the black arrow sketched below the three panels of Fig.~\ref{AL_HM_AIM}. 
Heuristically, one could imagine to start from the AIM case (where ${U_{MIT} \!= \! +\! \infty}$). Then, by progressively moving the value of $U_{MIT}$  first towards lower finite values and, subsequently, down to $0$, one could qualitatively ``reconstruct" the other two cases by squeezing the corresponding $\boldsymbol{\Gamma}_c^\infty$-lines against the corresponding accumulation points.

On a more formal level, we note that the presence or the absence of an accumulation point of vertex divergence lines, as those discussed here, might play an important role in the way vertex divergences may affect the (${T\! = \! 0}$)-physical properties of strongly correlated electron systems, such as, e.g.~the validity (or the violation) of the Luttinger theorem \cite{Fabrizio2022}.
Further, the presence of a increasing number of vertex divergences when approaching the Mott MIT at $T\! = \! 0$ from the correlated PM-side appears to provide a key to reconcile the analytical derivation of \cite{Kehrhein1998} with the numerical evidence \cite{Karski2005} of the occurrence of a smooth transition in DMFT calculations. Indeed, in \cite{Kehrhein1998} the impossibility of observing such a smooth Mott MIT at zero temperature was demonstrated by assuming the convergence of the (self-consistent) skeleton expansion for all $U\!<\!U_{c}(T\! = \! 0)$, an assumption which our study (see also the Appendix of \cite{Schaefer2016}) demonstrated to break down for $U$ much smaller than $U_{c}(T \! = \! 0)$.

\FloatBarrier
\section{\label{Conclusion}Conclusion and Outlook}

In this work, we have investigated the relation between a characteristic aspect of the breakdown of the self-consistent perturbation expansion \cite{Gunnarsson2017}, namely the occurrence of divergences of irreducible vertex  functions in the charge channel \cite{Schaefer2013,Schaefer2016} and the Mott-Hubbard metal-to-insulator transition. To this aim, by performing DMFT calculations on the two-particle level for the half-filled Hubbard model on a Bethe lattice, we have systematically studied the occurrence of irreducible vertex divergences in the coexistence region adjacent to the MIT. Our results demonstrate how the shape and the number of irreducible vertex divergence lines ($\boldsymbol{\Gamma}_{c}^\infty$-lines) in the coexistence region is significantly different in the PM and the PI solutions, with an abrupt jump across the thermodynamic first-order transition line, reflecting the different degree of suppression of on-site charge fluctuations \cite{Gunnarsson2016,Gunnarsson2017,Chalupa2021,Adler2022} in the two phases. Further, we could show that, in spite the evident backward bending displayed by the $\boldsymbol{\Gamma}_{c}^\infty$-lines in the correlated PM phase, the number of divergences crossed by approaching  the ${T\!=\!0}$ MIT at ${U\!=\!U_{c2}(T\!=\!0)}$ is diverging, making the location of the MIT itself an accumulation point for $\boldsymbol{\Gamma}_{c}^\infty$-lines. This finding establishes a clear connection between the irreducible-vertex divergences and the occurrence of the MIT, clarifying the difference with cases where vertex divergences appear in systems where no MIT takes place, and substantiating the interpretation of the Mott transition as highly-nonperturbative phenomenon in a more precise context.

Beyond the possible algorithmic implications of our results, especially relevant for Feynman diagrammatic approaches  to treat intermediate-to-strong coupling regimes \cite{Rohringer2018,DelRe2019}, it would be interesting, in the future, to extend our study by including the effects on non-local correlations, e.g.~by means of cluster extensions \cite{Maier2005} of DMFT. In particular, one could investigate, whether the shift towards lower $U$ values of the Mott MIT, found \cite{Maier2005,Park2011} by including spatial correlations of progressively larger size in the two-dimensional Hubbard model on a square lattice as well as the change of slope in the associated transition line $U_c(T)$,  would be accompanied by a corresponding shift of the accumulation point of the $\boldsymbol{\Gamma}_{c}^\infty$-lines as well as by a corresponding change of their bending. This piece of information would be of particular importance for precisely characterizing the nonperturbative nature of electronic correlations in two-dimensional systems, for which the location of the MIT, in absence of geometrical frustration, may be shifted \cite{Schaefer2015} down to ${U\!=\!0^+}$ in the low-temperature limit.

\begin{acknowledgments}
We thank S.~Andergassen, S.~Ciuchi, P.~Chalupa-Gantner, H.~E{\ss}l, D.~Fus, G.~Rohringer, G.~Sangiovanni and T. Sch\"afer for very insightful discussions and P.~Chalupa-Gantner also for carefully reading the manuscript. We acknowledge financial support from the Austrian Science Fund (FWF), within the project I-5487  (M.R. and A.T.) as well  as  project I 5868 (Project P01, part of the FOR 5249 [QUAST] of the German Science Foundation, DFG) (S.A.).
Matthias Reitner acknowledges support as a recipient of a DOC fellowship of the Austrian Academy of Sciences.
Calculations have been performed on the Vienna Scientific Cluster (VSC-4).
\end{acknowledgments}

\appendix
\section{\label{Workflow}DMFT-calculations of the two-particle Green's function $G^{(2)}$}
As mentioned in Sec.~II~D, we use a continuous-time quantum Monte Carlo solver (CT-QMC)  in the hybridization expansion (CT-HYB) \cite{Gull2011} of the \textit{w2dynamics} package \cite{Wallerberger2019,Kowalski2019} for the DMFT calculations of the two-particle Green's function $G^{(2)}$ of the auxiliary impurity model  and performed our calculations on the Vienna scientific cluster (VSC-4). A data set containing all numerical data and plot scripts used to generate the figures of this publication is publicly available on the TU Wien Research Data repository \cite{dataset}.

The detailed calculation process for the two-particle Green's function for a specific phase point in $U$, $T$ is the following: 
\begin{enumerate}
	\item Calculations of the one-particle Green's function $G^{(1)}$ with a comparable low number of measurements (e.g.~${\sim\!10^5}$) at each DMFT iteration, to converge the DMFT self-consistence cycle. The convergence is verified by tracking the behavior of $G^{(1)}$ for the first Matsubara frequency and the double occupancy.
	\item The first computations are followed by a calculation with a high number of measurements (e.g.~${\sim\!10^7}$) and a small number of DMFT iterations to obtain $G^{(1)}$ as function  of Matsubara frequencies with a satisfactory error to noise resolution as starting point for a two-particle calculation. 
	\item One DMFT iteration including a calculation of $G^{(2)}$ with a high number of measurements (e.g.~${\sim\!10^7}$) using a frequency box size of ${100\!\times\!100}$ fermionic frequencies for high temperatures and a box size up to ${200\!\times\!200}$ frequencies for our lowest temperatures. Thereby, we ensure that the number of  negative eigenvalues, originating from lower frequency structure of $G^{(2)}$ (see e.g. \cite{Kunes2011,Rohringer2012,Tagliavini2018,Wentzell2020}), is not effected by a further increase of the frequency box size. Each individual computation required an average of $\sim$10000 CPU hours.
\end{enumerate}

\subsection{\label{sampling}Sampling methods}
Within the CT-HYB calculations of the \textit{w2dynamics} package, different sampling methods for the computation of one- and two-particle Green's function can be used. For most of the parameter regimes we have used partition function sampling with the superstate-sampling method \cite{Kowalski2019} (segment sampling \cite{Werner2006} yielded no computational advantage). 

\begin{figure}[t]
\includegraphics[width=\linewidth]{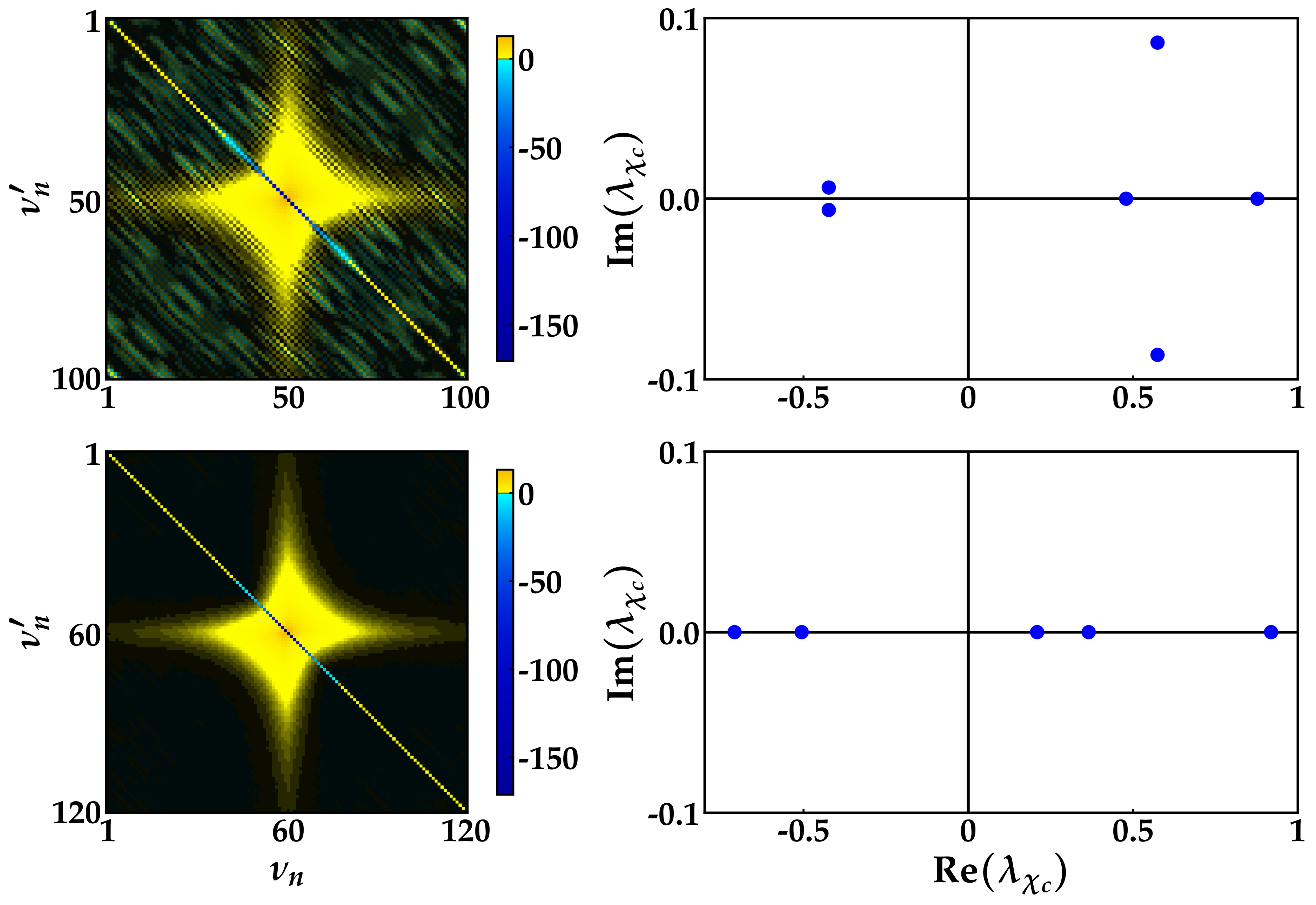}
\caption{\label{fig:artifacts}Left column: real part of the generalized charge susceptibility  $\boldsymbol{\chi}^{\nu \nu^{\prime}}_c$ for ${\beta\!=\!60}$ and ${U\!=\!2.46}$ for a PI solution. Right column: corresponding eigenvalues $\lambda_{\chi_c}$ around zero.
Upper row is calculated with state sampling, lower row with  worm sampling.}
\end{figure}

However, for the insulating solution in the coexistence region of our model and especially at low temperatures, partition function sampling runs into problems. This manifests in numerical artifacts, which occur as diagonal stripes in the $\boldsymbol{\chi}_c$-matrix causing complex pairs of the eigenvalues (see Fig.~\ref{fig:artifacts}), which are however prohibited by the particle-hole and $\operatorname{SU}(2)$-symmetry of the model considered ($\boldsymbol{\chi}_c$ must be a real bisymmetric matrix with only real eigenvalues \cite{Springer2020}), and therefore result in unusable data. This is presumably caused by the suppressed hybridization function $\Delta(\tau)$ in the insulating phase, resulting in a Monte Carlo estimator with high variance, stemming from functional derivatives of the partition function w.r.t.~the hybridization function \cite{Gunacker2015}. To overcome these numerical difficulties we used the worm-sampling method of \textit{w2dynamics} \cite{Wallerberger2019}, which uses Monte Carlo sampling in both the partition function and the Green's function space. Since, the worm-sampling method is numerically more expensive than superstate-sampling, we used worm-sampling only for lower temperatures in the insulating phase (${\beta\!>\!60}$).

\section{\label{Approx_of_lines}Approximation of vertex divergence lines}
Our numerical two-particle calculations for the determination of $N_{\lambda<0}$ have been performed for a finite set of $T$, $U$ parameters. To extract the points in the phase space, where the vertex divergence occurs, i.e. ${\lambda\!=\!0}$, from our data, we compared different approaches which are detailed in the following subsections. For our results in Sec.~III we used the polynomial fits described in Appendix \ref{Polynomial_approx}. For both methods the symmetry of the eigenvectors was not taken into account. Hence, a possible crossing of divergence lines \cite{Chalupa2018} was not investigated. Instead, the line ordering found in Refs.~\cite{Schaefer2016, Springer2020} is assumed. 

\subsection{\label{Eigenvalue_approx}Approximation via eigenvalues of $\boldmath{\chi}_c$}

\begin{figure}[t]
\includegraphics[width=0.59\linewidth]{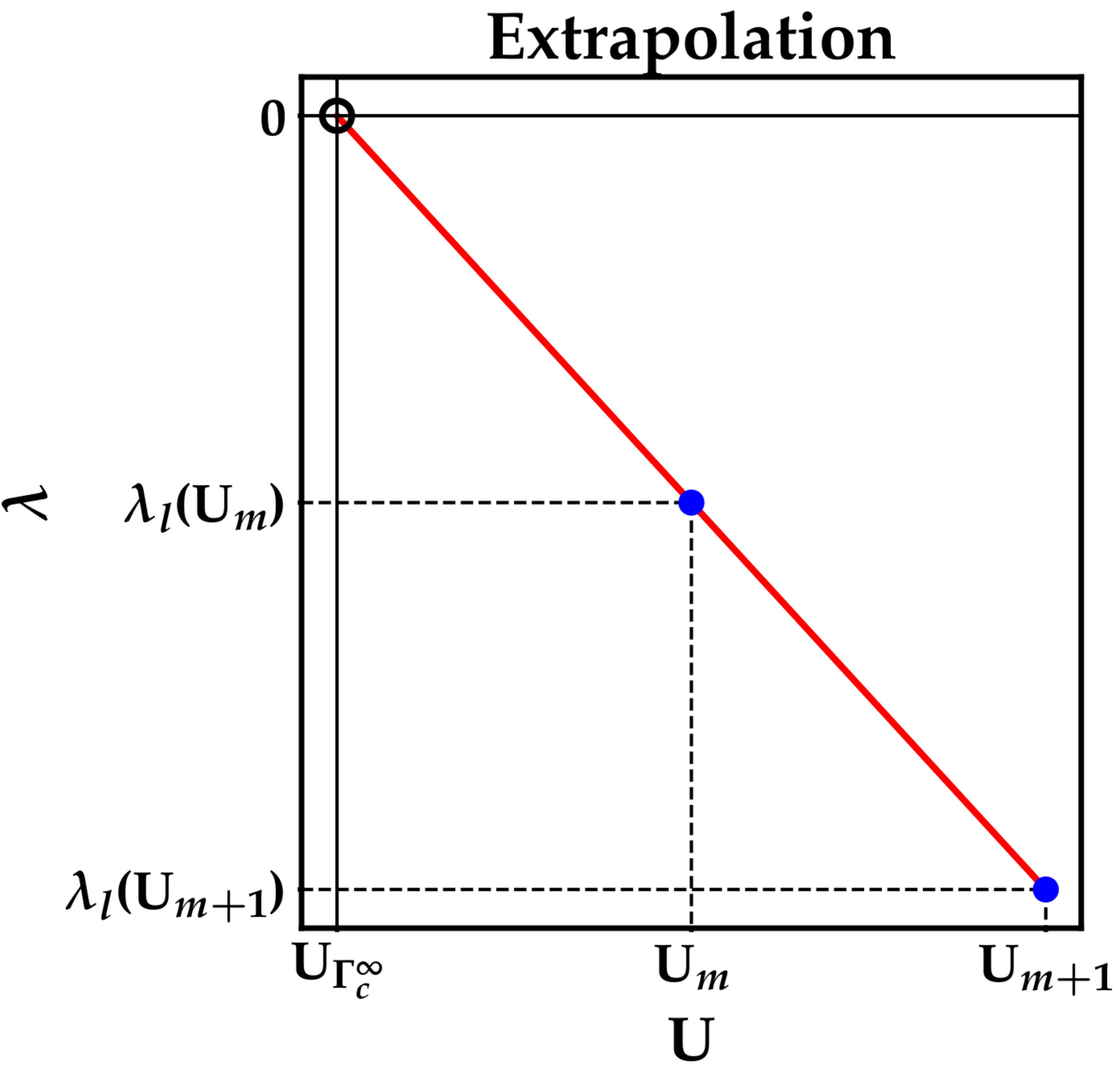}
\caption{\label{fig:extra-interpolation}Sketch of the extrapolation scheme to obtain the interaction value ${U\!=\!U_{\boldsymbol{\Gamma}_c^\infty}}$ of the zero-crossing of the  eigenvalue $\lambda_l$ of $\chi_c^{\nu\nu^\prime}$ from two-particle DMFT calculations at different $U_m$ for a fixed temperature $T$.}
\end{figure}

For fixed temperature $T$ scans with two-particle calculations close to the divergence, one can use a straight forward approach to extract the approximate interaction value $U_{\boldsymbol{\Gamma}_c^\infty}$, where ${\lambda\!=\!0}$ and a vertex divergence occurs: Linearly extrapolating from the first two data points, where ${\lambda\!<\!0}$ is already negative. $U_{\boldsymbol{\Gamma}_c^\infty}$ is obtained, where the extrapolation equals zero. This is schematically presented in Fig.~\ref{fig:extra-interpolation}. The corresponding $\boldsymbol{\Gamma}_{c}^\infty$-line of ${\lambda_l\!=\!0}$ for $U,T$ is approximated by connecting the obtained results for different $T$. The results for the $\boldsymbol{\Gamma}_{c}^\infty$-lines are presented in Fig.~\ref{fig:approx_comparison}, where the left panel corresponds to the extrapolation approach.

\subsection{\label{Polynomial_approx}Approximation via number $N_{\lambda<0}$ of negative eigenvalues of $\boldmath{\chi}_c$}
\begin{figure}[h]
\includegraphics[width=\linewidth]{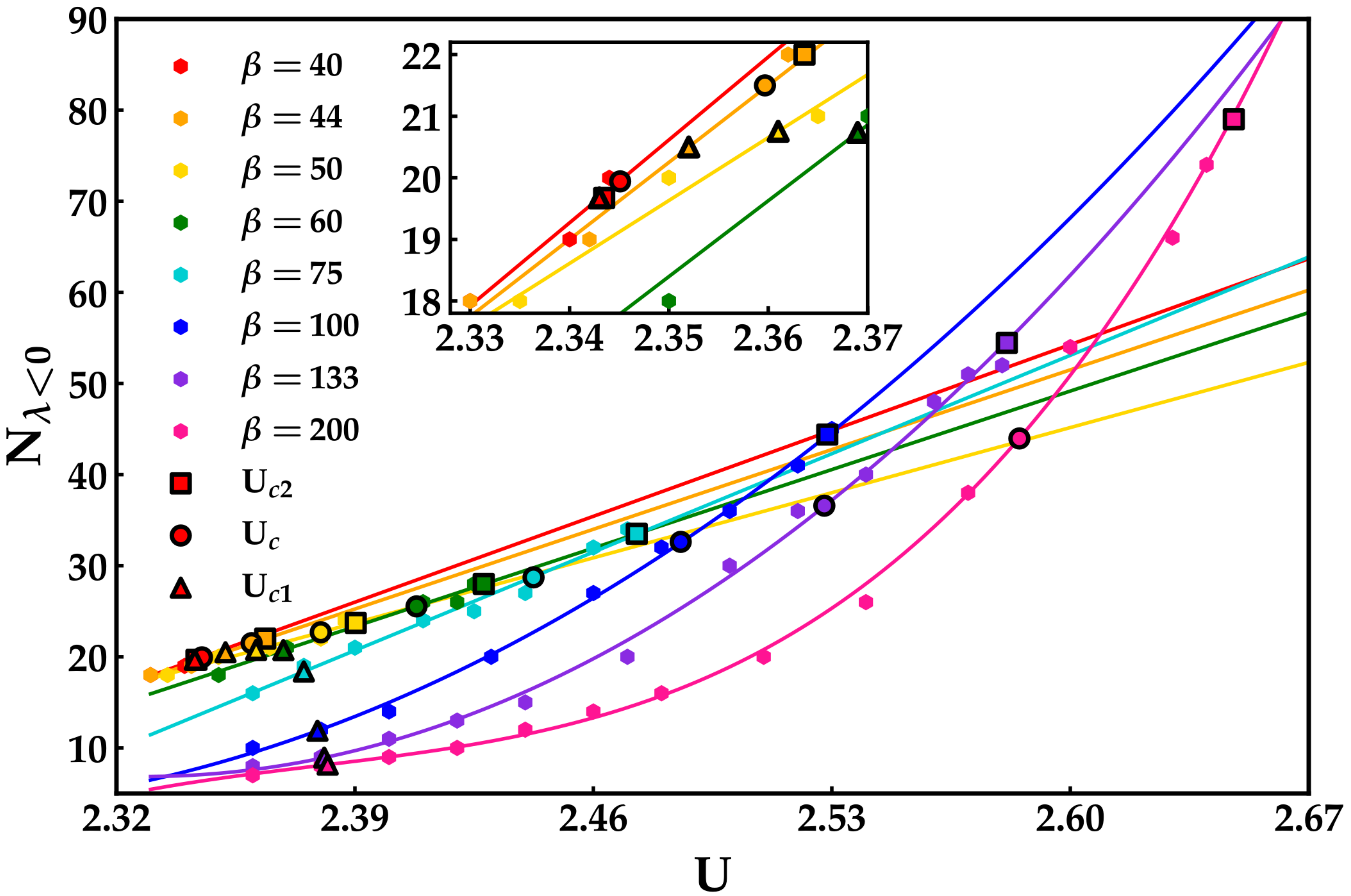}
\caption{\label{fig:polynomial_fits}Polynomial fits of $N_{\lambda<0}(U)$ for several temperatures. Black markers indicate $U_{c1}(T)$ (triangles), $U_c(T)$ from \cite{Blumer2002} (circles), and $U_{c2}(T)$ (squares) for the temperatures of the fits. The inset shows a zoom for low $\beta$ and $U$.}
\end{figure}

\begin{figure*}[t]
\includegraphics[width=0.8\linewidth]{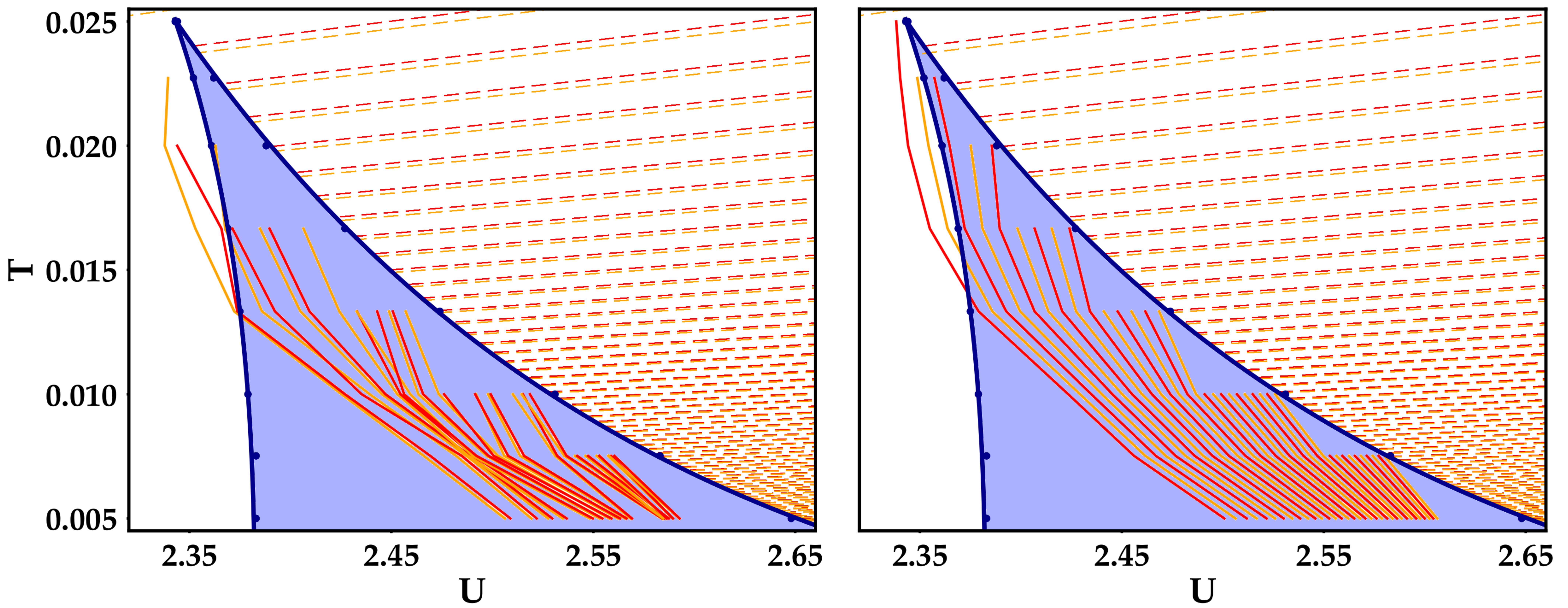}
\caption{\label{fig:approx_comparison}Phase diagrams with approximated $\boldsymbol{\Gamma}_c^\infty$-lines for the PM solution in the coexistence region, according to the approximations of \ref{Eigenvalue_approx} and \ref{Polynomial_approx} for several fixed $T$: (left) extrapolation of the eigenvalues of $\chi_c^{\nu\nu^\prime}$ and (right) polynomial fits of $N_{\lambda<0}$ along $U$. While the precise position of the $\boldsymbol{\Gamma}_c^\infty$-lines in $T,U$ differs between the different approximation schemes, the qualitative behavior of $N_{\lambda<0}(U,T)$ does not dependent on them.}
\end{figure*}
Alternatively to the previously described extrapolation scheme to approximate the $\boldsymbol{\Gamma}_c^\infty$-lines, $N_{\lambda<0}$ may be fitted by polynomial functions $p(U)$ for fixed $T$ by using the least square method. This approach, using the total number of negative eigenvalues, will in general yield less accurate positions of the divergence lines as compared to the approach of Appendix \ref{Eigenvalue_approx}. However, this approach takes the non-linear behavior of $N_{\lambda<0}$ along the temperature scans into account. The fits of the metallic coexistence region are shown in Fig.~\ref{fig:polynomial_fits}. For those fits we used $p(U)$ with different polynomial degrees. The polynomial degree thereby increases with decreasing $T$ (increasing $\beta$), due to the rapid non-linear increase of $N_{\lambda<0}$: 
${p(U)\!\propto\!U}$ for ${\beta\!\in\!\{40,44,50,60,75\}}$, ${p(U)\!\propto\!U^2}$ for ${\beta\!\in\!\{100,133\}}$ and ${p(U)\!\propto\!U^3}$ for ${\beta=200}$.  The $\boldsymbol{\Gamma}_{c}^\infty$-lines are approximated by connecting the points in $T$, $U$ with the same value of $N_{\lambda<0}$. The results for these lines are shown in the right panel of Fig.~\ref{fig:approx_comparison}.

Let us stress that, although, the precise position of the $\boldsymbol{\Gamma}_c^\infty$-lines in $T,U$ differs between the different approximation schemes, the overall behavior of $N_{\lambda<0}(U,T)$ and its asymptotic for ${T\!\to\!0}$ does not dependent on them.
Eventually, we note that, to generate $N_{\lambda<0}$ as function of $U$, $T$ for the parameter space, which is shown as color scale plots in Sec.~\ref{Results}, an additional linear interpolation between different temperatures $T$ has been used.

\section{\label{effectiveHA}Effective Hubbard atom like description of the insulating Hubbard model}

To approximately describe the behavior the divergence lines of the Mott insulating phase of the Hubbard model with Eq.~(\ref{HMPI}) and to account for the correct limit of ${U_{eff}\!\to\!\infty}$, we can construct a simple function \begin{equation}
    \eta(U)=\frac{U_{c1}^{T=0}}{U}(\eta_0-1)+1\ \text{,} \label{eq:eta}
\end{equation} where ${\eta_0\!=\!1.19\!\pm\!0.03\!\approx\!U^{T=0}_{c1}/2}$ is the mean shift between $\boldsymbol{\Gamma}_{c}^\infty$-lines of the HM and the HA within the coexistence region. In Eq.~(\ref{eq:eta}) we interpolate between $\eta_0$ at the beginning of the Mott insulating phase and ${\eta\!\rightarrow\!1}$ at ${U\!\rightarrow\!\infty}$ to account for the asymptotic behavior of the $\boldsymbol{\Gamma}_{c}^\infty$-lines. Hence, for $U_{eff}$ in $\mathcal{H}_{PI}$ of Eq.~(\ref{HMPI}) we get
\begin{align}
    U_{eff}&=\frac{U}{\eta(U)}=\frac{U^2}{U+U_{c1}^{T=0}(\eta_0-1)}=\nonumber\\[5pt]
    &\overset{2\eta_0\sim U_{c1}^{T=0}}{=}\frac{U^2}{U+{U_{c1}^{T=0}}(\frac{1}{2}U_{c1}^{T=0}-1)}
\end{align}
The results for the temperature behavior of $N_{\lambda<0}$ of the PI solution in Sec.~\ref{Accumulation} suggest to introduce another way to evaluate $\eta_0$, by comparing the slopes $\alpha_i$ of $N_{\lambda<0}(\beta)$ in Fig.~\ref{fig:Nvert_beta_ins} of Sec.~\ref{Insulating_results} between the HA and the HM at the same $U$. We can estimate the parameter $\eta_0$  by ${\eta_0^{\text{slope}}\!=\! \alpha_{HA}/\alpha_{HM}}$. The resulting values are listed in Tab.~\ref{tab:eta0}.

\begin{table}[h]
\centering
\begin{tabular}{|l|ccc|}
\hline
              &\multicolumn{1}{c|}{\ \ $U_{c1}(T)$\ \ }&\multicolumn{1}{c|}{\ \ $U_{c}(T)$\ \ } &\ \ $U_{c2}(T)$\ \  \\ \hline
\ \ $\eta_0^{\text{slopes}}$\ \         & \multicolumn{1}{c|}{1.22}     & \multicolumn{1}{c|}{1.16}    & 1.15     \\ \hline
\end{tabular}
\caption{Parameter $\eta_0$ calculated from the quotient of the slopes of $N_{\lambda<0}$ along the transition lines $U_{c1}(T)$,  $U_{c}(T)$ and $U_{c2}(T)$ in Fig.~\ref{fig:Nvert_beta_ins} for the effective HA description of the HM in the insulating phase.}
\label{tab:eta0}
\end{table}
In Tab.~\ref{tab:Uefftest} we compare ${n_{HA}\!=\!N_{\lambda<0}}$, calculated from $U_{eff}(\eta_0)$ for two different $\eta_0$, with the exact ${n_{HM}\!=\!N_{\lambda<0}}$ of the HM for two $U$ values at ${\beta\!=\!40}$: ${\eta_0^{\text{slope}}\!=\!1.22}$ at $U_{c1}(T)$ (the first $U$ value where a Mott insulating phase is possible) and ${\eta_0\!=\!1.19}$, from our considerations in Sec.~\ref{Insulating_results}. We see that $n_{HA}$  for $\eta_0^{slope}$ provides us a particularly good approximation for the Mott insulating phase of the Hubbard model.

\begin{table}[h]
\centering
\begin{tabular}{|l|l|c|c|} 
\hline
\multirow{2}{*}{$\beta=40$} & \multirow{2}{*}{$n_{HM}$} & \multicolumn{2}{c|}{$n_{HA}$ at~$U_{eff}$}  \\ 
\cline{3-4}
                            &                           & $\eta_0$ & $\eta_0^{\text{slope}}$                \\ 
\hline
$U=3.1$                     & \multicolumn{1}{c|}{31}   &\ \  30 \ \  & 30                               \\ 
\hline
$U=3.5$                     & \multicolumn{1}{c|}{34}   & 36       & 34                               \\
\hline
\end{tabular}
\caption{Test calculation results for the number of crossed vertex divergence lines and corresponding values according to the Hubbard atom with effective interaction $U_{eff}$ for ${\eta_0\!=\!1.19}$ and ${\eta_0^{\text{slope}}\!=\!1.22}$.}
\label{tab:Uefftest}
\end{table}

\section{\label{intermediate_lines}Calculation of intermediate lines}
The dashed dotted lines in the left panel of Fig.~\ref{fig:Nvert_beta_met} are intermediate lines interpolating between the $U_{c1}(T)$, $U_c(T)$, and $U_{c2}(T)$ lines according to
\begin{align}
    &U^{int}_{1}(T) =\frac{1}{2}\,\bigl[U_{c1}(T)+U_{c}(T)\bigr]\label{Ui1}\\
    &U^{int}_{2}(T) =\frac{1}{6}\,\bigl[U_{c1}(T)+5\,U_{c}(T)\bigr]\\
    &U^{int}_{3}(T) =\frac{1}{4}\,\bigl[3\,U_{c}(T)+U_{c2}(T)\bigr]\\
    &U^{int}_{4}(T) =\frac{1}{2}\,\bigl[U_{c}(T)+U_{c2}(T)\bigr]\\
    &U^{int}_{5}(T) =\frac{1}{4}\,\bigl[U_{c}(T)+3\,U_{c2}(T)\bigr]\label{Ui5}.
\end{align}
We used the polynomial fits (Fig.~\ref{fig:polynomial_fits}) to evaluate the corresponding $N_{\lambda<0}$ along those lines at inverse temperatures ${\beta\!=\!\{40,44,50,60,75,100,133,200\}}$ and applied the same fitting routine as along $U_{c1}(T)$, $U_c(T)$, and $U_{c2}(T)$ to generate the corresponding dashed dotted lines in the right panel.

\bibliography{library}

\begin{thebibliography}{82}
\bibitem [{DMF()}]{DMFT_boarder}%
  \BibitemOpen
  \href@noop {} {}\bibinfo {note} {The specific borders ($U_{c1,c2}(T)$) of the
  coexistence region have been also explicitly recalculated for this work,
  finding an essential agreement with the existing results \cite{Blumer2002},
  while for $U_c(T)$, which is determined by comparing the free energies of the
  PM and PI phases, literature results \cite{Blumer2002} have been
  used.}\BibitemShut {Stop}%
\bibitem [{DOS()}]{DOS_choice}%
  \BibitemOpen
  \href@noop {} {}\bibinfo {note} {As for many other properties, at the level
  of a DMFT analysis, the specific choice of the DOS, e.g., the one of a square
  lattice or the semicircular one of the Bethe lattice, does not have a
  qualitative impact on the final results. Furthemore, quantitative
  discrepancies can be considerably minimized, by choosing, as we do here,
  DOSes with equal standard deviations.}\BibitemShut {Stop}%
\bibitem [{MIT()}]{MIT_eta}%
  \BibitemOpen
  \href@noop {} {}\bibinfo {note} {Note that in the regime of the MIT
  ${\eta\!=\!1.19\!\pm\!0.03}$ is roughly equal to half of the interaction
  value of $U_{c1}$ at ${T\!=\!0}$
  (${\frac{1}{2}U_{c1}^{T=0}\!\approx\!1.19}$), which is the starting point of
  a metastable Mott insulating phase. For more details about the effective HA
  like description of the PI phase, see Appendix
  \ref{effectiveHA}.}\BibitemShut {Stop}%
\bibitem [{fit()}]{fitting_function}%
  \BibitemOpen
  \href@noop {} {}\bibinfo {note} {The applied fitting function used to fit the
  high $\beta$ range (${\beta\!\geq\!75}$) was
  ${N_{\lambda<0}^{U_{c1}}(\beta)\!=\!a_1\!+\!b_1(\beta\!-\!\beta_0)^{-1}}$,
  which behaves as
  ${N_{\lambda<0}^{U_{c1}}(\beta)\!\cong\!a_1\!+\!b_1\beta^{-1}}$ for
  ${\beta\!\rightarrow\!\infty}$}\BibitemShut {NoStop}%
\bibitem [{sim()}]{simultaneous_divergencies}%
  \BibitemOpen
  \href@noop {} {}\bibinfo {note} {We do not consider here, the obvious case of
  second-order phase transitions, characterized by simultaneous divergences of
  reducible and irreducible vertices.}\BibitemShut {Stop}%
\end{thebibliography}

\end{document}